\documentclass[3p, twocolumn]{elsarticle}

\usepackage{amsmath}
\usepackage{graphicx}
\usepackage{multirow}
\usepackage{array}
\usepackage[table]{xcolor}
\usepackage{tabularx}
\usepackage{subcaption}
\usepackage{hyperref}
\usepackage{etex}
\usepackage[switch]{lineno}
\usepackage{pifont}
\usepackage{makecell}

\usepackage{xtab,booktabs}

\newcommand{\abc}[1]{\textcolor{black}{#1}}
\newcommand{\defg}[1]{\textcolor{black}{#1}}

\journal{Journal of Systems and Software}
\begin{document}

\begin{frontmatter}
\title{
Preprint of the paper accepted at Journal of Systems and Software.\\
The fianl version of the paper can be accessed here: \\ {\normalsize
$-$ \url{https://doi.org/10.1016/j.jss.2022.111479} OR\\ 
$-$ \url{https://www.sciencedirect.com/science/article/abs/pii/S0164121222001637}} \\ 
----------------------------------------------------------------------------------------\\
-----------------------------------------------------------------------------------------\\
On the Benefits And Problems Related to Using \\Definition of Done --- A Survey Study }

\author[put]{Sylwia Kopczy\'{n}ska\corref{cor}}

\ead{skopczynska@cs.put.poznan.pl}        

\author[put]{Mirosław Ochodek}

\author[put]{Jakub Piechowiak}

\author[put]{Jerzy Nawrocki}

%\ead{www.elsevier.com}

\cortext[cor]{Corresponding author}

\address[put]{Poznan University of Technology \\
			  Faculty of Computing, Institute of Computing Science \\
              ul. Piotrowo 2, 60-965 Pozna\'{n}, Poland \\}

\begin{abstract}

\noindent\textbf{Context:} Definition of Done (DoD) is one of the fundamental concepts of Scrum. It expresses a shared view of a Scrum Team on what makes an increment of their product complete. DoDs are often defined as checklists with items being requirements towards software (e.g., quality requirements) or towards activities performed to make the increment shippable (e.g., code reviews, testing). Unfortunately, the knowledge about the usefulness of DoD is still very limited. 

\noindent\textbf{Objective:} The goal is to study what benefits using the DoD practice can bring to an agile project, what problems it may trigger, and how it is created and maintained.

\noindent\textbf{Method:} In the survey among members of agile software development projects, 137 practitioners from all over the globe shared their experience with us.

\noindent\textbf{Results:} 93\% of the respondents perceive DoD as at least valuable for their ventures. It helps them to make work items complete, assure product quality, and ensure the needed activities are executed. However, they indicated that every second project struggles with infeasible, incorrect, unavailable, or creeping DoD.

\noindent\textbf{Conclusions:} It follows from the study that DoD is important but not easy to use and more empirical studies are needed to identify best practices in this area.

\end{abstract}


\end{frontmatter}
 
\modulolinenumbers[5]
%\linenumbers

\section{Introduction}
Organizations are increasingly adopting agile methods. From the results of the PMI survey, it follows that over 70\% of organizations report using agile methods \cite{PMIPulse2017}, and Scrum is one of the most popular among them.

In Scrum, Developers are to deliver increment(s) of ``done'' product in each Sprint. It means that an increment must be complete and in usable condition (often referred to as working software) and meet the agreed Definition of Done (DoD). According to the Scrum Guide \cite{scrum-guide} ``\textit{the Definition of Done is a formal description of the state of the Increment when it meets the quality measures required for the product.}'' DoD might be standardized at the level of organization, or could be created separately for each product. Also, it usually evolves in time to include more stringent criteria as the team matures. 
Although the use of DoDs is characteristic to the Scrum framework, it is also recommended by other agile/lean frameworks like The Kanban Method \cite{theobald2019comparing}, and appears in several Agile Maturity Models (AMMs) \cite{nurdiani2019understanding}.

The most recent version of Scrum Guide \cite{scrum-guide}  states that a DoD ``\emph{creates transparency by providing everyone a shared understanding of what work was completed as part of the Increment.}'' Therefore, it is essential that a DoD is formulated in such a way that all team members understand it in the same way. \abc{Although there are no official guidelines on how to achieve this, most of the teams end up defining their DoDs as lists of ``items'', which need to be satisfied to claim that a given Sprint is ``done''. DoD items concern, e.g., source code quality, performing code review, testing activities, completing work items, or non-functional requirements regarding cross-cutting concerns, like security or performance. \cite{Silva2017,ALSAQAF201939}} 

The existing body of knowledge regarding the practical usage of DoD is very limited \cite{Silva2017}.  
%There were two main studies focusing on understanding how DoDs are defined and used. The first study by Silva et al. \cite{Silva2017} was a Systematic Literature Review that included papers published prior to 2017. The authors found and analyzed 8 examples of DoDs what allowed them to identify several categories of DoD items. In their follow-up study \cite{Silva2018}, they surveyed 20 professionals asking them about how DoDs are created and maintained in practice. Although these studies shed some light on how DoDs are used in agile projects, their empirical basis is limited. Thus, more research is needed to thoroughly analyze the phenomenon of DoD. 
Taking into account little awareness of the importance of working towards the shared understanding of what ``done'' means in some teams \cite{OConnor2010,paasivaara2018large}, it would be especially valuable to study the potential benefits and problems related to using DoD to bring new arguments to the discussion about how much we should invest in the preparation of DoDs.

The goal of this paper is to investigate how practitioners use Definition of Done (DoD) in software projects and product teams. In the further paragraphs, for simplicity, we will refer to both software projects and product teams as \emph{projects}\footnote{\abc{We investigate both software development initiatives planned for a certain period of time (projects) as well as those continuously developing some products (product teams).}}. The special focus of the paper is on the usefulness and the problems DoD might trigger. To achieve the goal, we conducted a survey among practitioners from all over the world. The main contributions of our study are as follows:
\begin{itemize}
    \item we investigated to what extent the practitioners perceive the DoD practice as valuable,
    
    \item we studied 19 (and identified two more) benefits of using DoDs by investigating their frequency of appearance in agile projects,
    
    \item we identified 19 problems that are triggered by lack of using DoD in a project,
    
    \item we studied the frequency and significance of 19 problems that agile teams might encounter while using DoDs,
    
    \item finally, we investigated \abc{what DoD items are}, and the process of creating and maintaining DoDs, focusing on the steps of that process, the roles that are involved, and the tools that are used.

\end{itemize}

\abc{This paper is organized as follows. First, we explain what DoD is in Section \ref{sec:background}. Then, in Section \ref{sec:relatedwork}, we discuss the related studies. Next, in Section \ref{sec:method},  we describe the design of the survey and discuss the validity threats. The demographic  information about our respondents is presented in Section \ref{sec:Demography}, while the results of the survey are presented in Section \ref{sec:results}. The implications of our study are discussed in Section \ref{sec:discussion}. Section \ref{sec:conclusions} concludes our findings.}

\section{Background}
\label{sec:background}
According to Scrum Guide~\cite{scrum-guide}, each Sprint is to deliver an Increment of a potentially releasable product which is usable and adheres to the Scrum Team's ``Definition of Done''. However, the guide does not provide specific guidelines on how DoD should look like. However, it draws attention to its high value, i.e., complying with DoD is required at the end of each Sprint to ensure transparency and assure product quality. Practitioners describe DoD as a ``social contract in agile teams'' that ``acts as a check before work is allowed to leave the Sprint'' \cite{power2014definition}.

However, the concept of DoD is not only used in Scrum but also in other agile approaches. For example, The Kanban Method recommends to define the workflow, that is the specific process each item-to-do needs to undergo with the defined steps (stages) and the policies specifying the requirements that need to be satisfied to let the item flow from one step to another. One type of such policies is DoD \cite{anderson2016kanban}.

%as one type of explicit policies \cite{anderson2016kanban}. The policies articulate and define the process that goes beyond the definition of the workflow. They need to be simple, well-defined, visible, always applied, and readily changeable \cite{anderson2016kanban}. And, thus, they are usually visible on Kanban boards at the top of each column. Each column represents a step of a software development process, and thus, DoD items state the constraints on the work item processed within a certain step.

\begin{figure}[h!]

\centering\includegraphics[width=0.9\columnwidth]{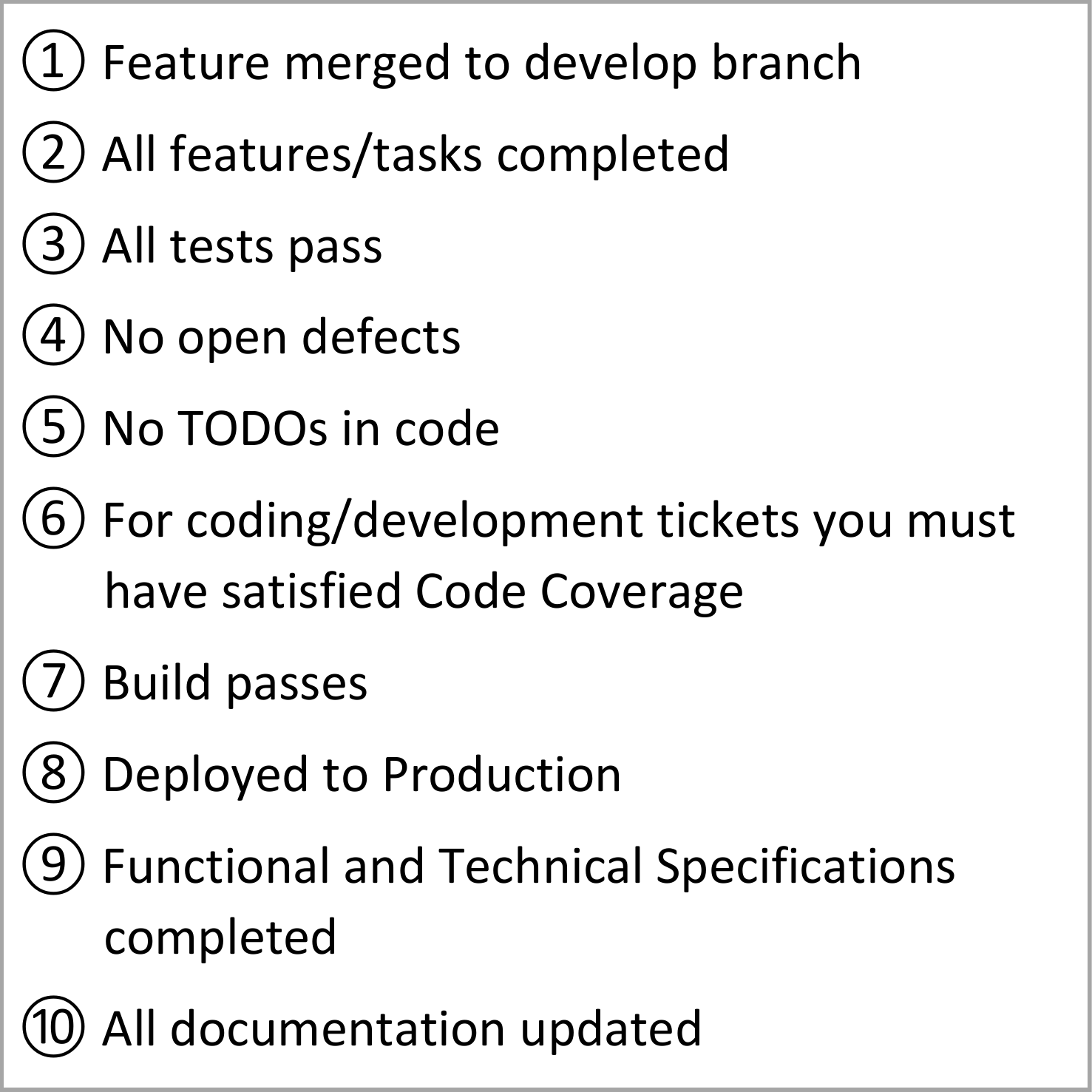}
\caption{Excerpts from DoDs presenting different types of DoD items.}
\label{fig:dodExamples}
\end{figure}

In practice, DoD frequently has a form of a checklist with requirements that need to be assessed to determine whether the work is ``done'', see examples in Figure~\ref{fig:dodExamples} (the examples \ding{192}-\ding{201} are taken from the DoD items the respondents shared with us in the survey). DoD items mainly focus on \cite{Silva2017, Madan2019, kopczynska2020importance, Saddington2012, ALSAQAF201939}: (1) business or functional requirements, i.e., require certain requirement(s) to be implemented or tested or verified, e.g., \ding{193} ; (2) quality aspects, e.g., concern the status of testing, the level of technical debt, the results of code reviews such as \ding{194}, \ding{195}, \ding{196}, \ding{197}, and (3) non-functional requirements, i.e., they state if the desired characteristics of a system are met, e.g., \ding{200}. They might concern different types of entities that appear in software development (e.g., test (\ding{194}), code (\ding{196}), defects (\ding{195}), documentation (\ding{200})) and be defined at multiple ``levels'', e.g., for a single task, for a user story/feature (\ding{192}), for all user stories/features (\ding{193}), for a whole increment (\ding{198}), for a release (\ding{199}).

DoD items are sometimes confused with Acceptance Criteria (AC). However, those two have slightly different goals. AC is a set of requirements that must be fulfilled for a given Product Backlog Item, so it can be accepted by Product Owner, customer, user, or other stakeholder \cite{Rubin2013}. Testing against AC produces a clear pass/fail result if the product can be accepted or not \cite{Povilaitis2014}. Thus, ACs are item-specific and orthogonal to DoD.  Viscardi \cite{Viscardi2013} states that DoD is a quality goal, while AC are functional or behavioral expectations. AC are not directly mentioned in the Scrum Guide, however they are an important and useful tool in agile projects. \defg{ Example ACs for an e-commerce site, which follow from the experience of the authors,} could be AC1:``\textit{Lists 1.000 products on a one page}'' or AC2: ``\textit{Sorts the listed products with respect to price and name in both ascending and descending order}'' while  DoD items could be DoDI1: ``\textit{Deployed to the production environment}'', DoDI2: ``\textit{Each feature was code reviewed by at least one dev who is not the author and the improvements are introduced}''.

While the notations of user stories or use cases are frequently used to document software requirements \cite{cohn2010succeeding}, there is no common format in which DoD items are specified. Frequently, DoD items are short statements or short sentences such as those presented in Figure~\ref{fig:dodExamples} or mentioned in the work of Silva et al. \cite{Silva2018}.

%Power2014 - "DoD and DoR are social contracts in agile teams, Together, they provide a boundary that stabilizes the team’s working environment, prevents 
% waste (time, delays, churn, working on the wrong things), remove impediments, and avoids the accumulation of technical debt and quality debt
% Definition of done acts as a check before work is allowed to leave the 
%Sprint. There is a saying we use from the agile community: “Let nothing into a Sprint 
%that is not Ready; let nothing out of a Sprint that is not Done”.

\section{Related work}
\label{sec:relatedwork}
The usage of DoD and its value were discussed in several papers describing lessons learned from industry projects. For example, according to Davis~\cite{Davis2013}, using DoD has a positive impact on reducing the number of defects and limiting technical debt. Power~\cite{power2014SocialContracts} claims that DoDs help with assessing the completion of work. Taipale~\cite{taipale2010Huitale} found that their software development with DoD results in the product of high quality and is a well-managed development process. Masood et al. \cite{masood2020agile} observed that a well-defined DoD supports self-assignment of tasks in agile teams. Kasauli et al. \cite{kasauli2020agile} proposed using DoD items at the level of user stories to foster requirements traceability in a project. Several other agile practitioners claim that DoD is an essential practice and recommend using it, e.g.,\cite{Saddington2012}, \cite{jakobsen2008mature}, \cite{cohn2010succeeding}. Although these papers mention some benefits of using DoD, they base their observations either on the authors' experience or on analyses of individual cases. Our study complements the previous findings by identifying and helping to understand the benefits of using DoDs in agile projects. We also scale up the existing research by surveying a large sample of projects. 

While the majority of authors discuss the advantages of creating and using DoD, only a few of them mention the problems that accompany that process. For instance, O'Connor \cite{OConnor2010} states that there might be some conflicts between business and developers on what ``done'' means, while Igaki et al. \cite{Igaki2014} emphasize the need of making sure that the DoD items are followed. Igaki et al. \cite{Igaki2014} and Alsaquaf et al. \cite{ALSAQAF201939} found that DoD might become too extensive and as a result negatively impact team velocity or make validation more complex.

Although these papers present valuable lessons learned from implementing and teaching agile methods, neither of them provides a comprehensive and well-evaluated set of problems related to using DoDs. With our study, we fill this knowledge gap. We use the identified problems and benefits as the basis, extend them with those from our knowledge and experience, and evaluate their importance among practitioners all over the world. 
We compared the related works mentioned in two previous paragraphs with each other and with our work in Table~\ref{tab:coparisionExp}.
\begin{table*}[!ht]
\caption{Comparison of related work concerning experience in agile software development that revealed benefits or problems concerning the usage of DoD}

{\footnotesize
	\renewcommand{\arraystretch}{1.1}
    {\setlength{\tabcolsep}{0.35em}
	\centering{
		\begin{tabular}{|p{2.3cm}| p{2.75cm} |p{2cm}|p{2.75cm}| p{2cm} | p{2.75cm} |}
		\toprule
		 & \textbf{Object of study} & \textbf{Type of study} & \textbf{Participants} & \textbf{Domain} & \textbf{Problems or Benefits} \\
		 \toprule
Power et al. \cite{power2014definition} & using Definition of Ready & case study & one company, overall opinion & networking &  identified a benefit \\	
\midrule
Taipale et al. \cite{taipale2010Huitale} & process used in Lean startup  & case study & one company, overall opinion & software development, retail &   identified a benefit \\	
\midrule
Massod et al. \cite{masood2020agile} & self-assignment to work & grounded theory & 23 companies, 42 practitioners & software development &   identified a benefit \\	
\midrule
Kasauli et al. \cite{kasauli2020agile} & challenges and strategies using agile approaches & case study & one company, two departments & automotive  &   identified a benefit \\	
\midrule
Saddington et al. \cite{Saddington2012} & scaling product ownership & case study & a program by the Department of Defense and US Air Force, 4 teams & military &   identified a benefit \\	
\midrule
Jakobsen and Johnson \cite{jakobsen2008mature} & combining CMMI with Scrum & case study & one company & software and systems company &   identified a benefit \\	
\midrule
Cohn \cite{cohn2010succeeding} & theory and experience about agile & no study, book & own experience & -- &   share possible benefits \\	
\midrule
O'Connor \cite{OConnor2010} & agile transformation & experience description & one company, one project & education & identified a problem   \\	
\midrule
Igaki \cite{Igaki2014} & ticket driven development for teaching Scrum & experiment with students & one company, one project & education & identified a problem   \\	
\midrule
Alsaqaf et al. \cite{ALSAQAF201939} & way developers treat quality requirements in agile & multi-case study, interviews & 17 agile practitioners from 6 organizations & public sector, government, commercial, banking, commercial navigation, health care, insurance, telecom & identified a problem   \\	
\midrule
Our Study & Practice of DoD & survey & 137 agile practitioners & over 20 different domains & analyzed problems and benefits and the process of using DoD    \\
\bottomrule
		\end{tabular}
	}
	}
}
\label{tab:coparisionExp}
\end{table*}

There are two works in which the focus was given to DoD. First, Silva et al.~\cite{Silva2017} conducted a Systematic Literature Review on the papers published prior to 2017. The review aimed at investigating what the done criteria are, what is the context of the teams using DoD, and what the characteristics of the studies investigating the topic of DoD are. They found 8 relevant studies and analyzed them. As a result, they characterized the items that are commonly present in DoDs. For instance, they observed that DoD items are often defined at four levels: story, Sprint, release, and project. They found that four primary studies mention the benefits of using DoD. Finally, based on the analysis of gaps in the literature regarding DoD, they concluded that ``\textit{there is a need for more and better empirical studies documenting and evaluating the use of the DoD in agile software development}.''
In their follow-up study, Silva et al.~\cite{Silva2018} conducted a survey on a sample of 20 respondents. The study provided some insights on how DoD is created and maintained, e.g., they learned that 80\% of respondents had their DoDs explicitly documented, the DoDs were mostly created by the team and evolved throughout the project. Our study scales-up and complements both of these studies, providing some new insights on DoD items, on how DoDs are created and maintained, and investigates the importance of the identified benefits and problems. The comparison of the studies with each other and with our study is presented in Table~\ref{tab:coparisionDo}.

\begin{table*}[!ht]
\caption{Comparison of related work concerning the practice of using DoD with each other and with our work.}
{\footnotesize
	\renewcommand{\arraystretch}{1.1}
    {\setlength{\tabcolsep}{0.35em}
	\centering{
		\begin{tabular}{|p{2.5cm}| p{4.3cm} |p{4.3cm}|p{4.3cm}| }
		\toprule
		 & \textbf{Silva et al. \cite{Silva2017}} & \textbf{Silva et al. \cite{Silva2018} }&\textbf{ Our work} \\
		 \toprule
		 \textbf{Type of study} &  SLR & survey & survey \\
		 \midrule
		 \thead[tl]{Research \\ Questions} & \makecell[tl]{1. What are done criteria? \\
2.  In which environment DoD \\was used? \\
3. What types of study tackle \\ DoD?} &  \makecell[tl]{1. What  done criteria are? \\
2. What process is used to \\ define DoD? \\
3. What levels of DoD are used?\\
4. Is DoD emergent during \\ a project?\\
5. Are DoDs explicitly \\ documented?\\
6. How are DoD criteria \\ assessed? \\
7. Effectiveness of DoD\\
8. Challenges in using DoD}
& \makecell[tl]{1. What are the benefits\\ of using DoD?\\
2. What are the problems\\ encountered while using DoD \\ and how they are mitigated?\\
3. Problems--Benefits\\ relationships\\
4. What DoD items are?\\
5. How is DoD established\\ and maintained?} \\

\midrule
\thead[tl]{Subjects/ \\Participants} & 8 primary studies & 20 agile practitioners from 16 countries & 137 agile practitioners from 45 countries\\
\midrule
		 \thead[tl]{\#DoD items \\ analyzed} & 62 & 22 & 143 \\
\midrule
		 \thead[tl]{Problems \\or Benefits} & Found the benefits mentioned in four primary studies & Asked respondents about benefits and challenges  &  Used the identified benefits and problems to survey practitioners from all over the world about perceived importance and identify any new items.\\
\midrule
		 {Practice of using DoD} & -- & Asked with open questions respondents about the definition, evolution, assessment, and if DoD is documented & Asked with open questions about the whole process of using and maintaining DoD, person responsible, and format.   \\
\midrule
		 {DoD item analysis} & Categorized DoD items and investigated the frequency & Listed the DoD items using the categorization previously proposed  & Analyzed DoDs using the previously proposed categorization \\

		 \bottomrule
		\end{tabular}
	}
	}
}
\label{tab:coparisionDo}
\end{table*}

%According to our up-to-date knowledge there are a few published attempts to systematically understand better what DoD is. First, in 2016, Oślizło~\cite{oslizlo2016} analyzed DoDs from 15 different companies. He found out that there are various categories of DoD items and there are several items that are very similar to each other and appear frequently. Silva et al.~\cite{Silva2017} conducted a literature review with papers published up to 2016 which allowed them to analyze 8 DoDs and identify several categories of items. 
%\sk{czy wspominianie o pracy pana Oslizlo bedzie naruszalo double-blind?}

%Finally, there are also some experience reports and blog posts that provide some examples of DoD items, e.g., ``project builds successfully'' \cite{Igaki2014}, ``code unit tested'' \cite{Igaki2014}, or  ``acceptance tests passed.'' \cite{OConnor2010}

%According to our up-to-date knowledge, there is no study that systematically analyze a broad range of DoD and of agile practitioners to better understand what DoD really is, what its value is, or what problems it might bring to the project.

%Although some of the above presented studies report positive effects and issues related to using DoDs, none of them attempts to systematically analyze the benefits or problems that agile teams may encounter while implementing this practice. In this study, we fill this knowledge gap by conducting a survey among the practitioners and asking them about the implications (both positive and negative) of adopting the DoD practice in their projects.

\section{Research methodology}
\label{sec:method}

\subsection{Research aims and questions}
\abc{Our goal is to study the usefulness of DoDs from two perspectives.  Firstly, we are interested in learning what benefits a DoD might bring to a project or product team.\footnote{Scrum and other agile approaches can be used both in software projects and in product teams. For simplicity, we will refer to both as \textit{projects} in this paper.} Secondly, we want to identify and evaluate the \abc{problems} associated with implementing the DoD practice. Finally, we would like to investigate whether there are any relationships between the problems encountered while using DoD and the benefits of using it.} \defg{The existence of such relationships would mean that when certain problems with DoD exist, some benefits are less/more likely to be observed. Thus, it might help to make decisions about problem-mitigation strategies.}
 
\abc{Based on these three goals, we formulate the following research questions:}
\begin{itemize}
    \setlength\itemsep{4pt}
    \item RQ1. \textit{What are the benefits of using DoD?}
    \item RQ2. \abc{\textit{a) What are the problems encountered while using DoD and b) how are they mitigated?}}
    
    \item \abc{RQ3. \textit{Are there any relationships between the problems and benefits while using DoD?}}
    
\end{itemize}

As follows from the literature review presented in Section~\ref{sec:relatedwork}, the body of knowledge regarding the process of creating and maintaining DoDs is very limited. Therefore, we formulate one more research question to broaden our understanding of that process:
\begin{itemize}
    \item \abc{RQ4.} \textit{How the DoD practice is implemented?} 
    \begin{itemize}
        \item \abc{\textit{a) How DoD is created and maintained?} }
        \item \abc{\textit{b) What are the DoD items? } }
    \end{itemize}
\end{itemize}
Additionally, answering this question would allow us to characterize the context in which the DoDs were created and used. Consequently, it might help us understand the origins of the benefits and problems.

\subsection{Research method}

We chose questionnaire-based Survey Research as our research method of choice. Since this method allows for collecting and analyzing large samples of projects in a cost-effective way, it gives us the possibility to draw an overall picture of what benefits and problems one might expect to encounter when adopting the DoD practice in a project. While designing our study, we followed the guidelines provided by Wohlin et al. \cite{Wohlin_etal_2012} and by Molléri et al. \cite{molleri2020empirically}. Our survey can be classified as a descriptive study, however, since the topic we investigated has not been deeply studied, it could also, to some degree, be treated as an exploratory study.

The questionnaire and the collected data are available as supplemental material for this paper.

\subsection{Population and sample representatives}
\label{sec:Population}

Our target population constitutes participants of agile software development projects and product teams that use DoD practice. 
%We would like to draw an overall picture of how the practice is used in agile projects and what are the consequences of using it. 
We assume that a representative sample of the target population shall have the following characteristics:
\begin{enumerate}

    \item Using the DoD practice --- the sample includes responses from the participants that used the DoD practice in the projects they refer to. 
    
    \item Year --- projects being referred to are contemporary agile projects. Referring to old projects would constitute a serious threat to internal and external validity since the usage of agile methods and their popularity has changed visibly over the last years \cite{hoda2018rise}.

    \item Context (country, domain, etc.) --- the distributions of different project context factors in the sample of projects should be similar to the corresponding distributions in the IT industry.
    
    \item Agile method --- we assume that the DoD practice is mainly used by the teams working according to the Scrum guidelines (since the DoD practice originates from Scrum). Therefore, we assume that our target population is dominated by Scrum teams. 
    
\end{enumerate}

We are going to evaluate the representatives of the sample based on the above given characteristics. The first two characteristics (1 and 2) are filtering criteria that should be included in the survey instrument. The two remaining characteristics (3 and 4) are more difficult to evaluate since we do not have accurate and precise information on how often given context factors appear in IT projects. Therefore, we are going to base our evaluation on the available data from surveys concerning agile methods \cite{versionone2021, ochodek2018perceived, Matharu2015} and the characteristics of the large sample of IT projects collected in the ISBSG database and reported by Hill \cite{hill2011practical}. In particular, we are going to take into account the geographic dispersion of the respondents/projects and the agile methods that are reported in the surveys, while the ISBSG database allow us to assess the representatives of different domains and types of applications in the sample.

%Our target population constituted participants of Agile software development projects and product teams. We would like to draw an overall picture of how the practice of DoD is used in agile projects. Therefore, we \emph{do not}  aim at any specific domain, application type, agile methodology, culture, country, etc. Our goal is to survey a group of respondents coming from different project environments. Ideally, the distribution of different context attributes in our study would be similar to how common they are in the IT industry. For instance, multiple studies reported that Scrum is currently a dominating Agile method (e.g., \cite{ochodek2018perceived, Matharu2015, hoda2018rise}). Therefore, we  would expect to see this trend reflected in the studied sample. 

\subsection{Survey instrument}

We designed an online questionnaire divided into four parts and implemented it using the Survey Monkey platform. 

The first part consisted of three pages: a welcome page (presenting the goal of the study and providing contact information to the researchers conducting the study), a page providing some basic information about the concept of DoD (as one of the means to ensure construct validity), and a page asking the respondents to focus on one of the projects or product teams they participated in which a DoD was used. We based the description of DoD on the Scrum Guide~\cite{scrum-guide}. Also, we asked about the date of the last participation of the respondent in a project in which a DoD was used. It had two purposes: (1) we wanted each respondent to focus on an individual project, and (2) we wanted to verify whether the responses regard new developments.  

The second part of the questionnaire regarded the value of using a DoD in a project. First,  we asked the respondents to evaluate how valuable was using a DoD in their project (in general) by using a five-point Likert scale. Second, there was a list of 19 potential benefits to be assessed individually. The list of benefits resulted from our literature review and was extended by a few potential benefits that we brainstormed. Also, the respondents could select the option ``other'' to provide their own examples of benefits. For each benefit, the respondent was asked if the benefit to their project is the result of using a DoD. They responded using a five-point ordinal scale (``Definitely YES'', ``YES'', ``Neither YES nor NO'', ``NO'', ``Definitely NO''). We also allowed for the answer ``I don't know'' to avoid introducing bias by forcing respondents to provide answers (e.g., if the respondents lacked the knowledge to answer the question reliably or did not want to express their opinion).

The third part of the questionnaire contained a series of questions concerning problems with using DoDs. The first two questions asked the respondent to assess the negative impact of the possible problems in their projects concerning management and DoD items, respectively. The list of problems had two sources. We had extracted the problems mentioned in the literature. Next, since DoD items are requirements, we extended the set with the common problems by specifying requirements from Wiegers~\cite{Wiegers2013}.
For each problem, the respondents could select one of the following answers: ``Problem did not appear in the project'', ``Problem had a positive impact'' (e.g., there was a problem with using DoD but its presence triggered some corrective actions that were beneficial for the project) or assess the significance of the negative impact of the problem if it had appeared using a five-point Likert scale (``Definitely significant'', ``Rather significant'', ``Neither significant nor NOT significant'', ``Rather NOT significant'', ``Definitely NOT significant'').  Again, we also allowed for the answers ``I don't know'' and ``Other''. Finally, we asked the respondents how the problems with using a DoD were solved (an open question).

In the fourth part, we placed open questions concerning the process of creating and maintaining DoDs (the process, the roles involved, the frequency of updating the DoD, and the ways of publishing it). 

The fifth part contained 11 questions asking about demographic information. We used them to characterize the sample of respondents.
%and to identify potential relationships between the demographic information and the opinions expressed in the previous parts. <- tylko jesli zrobimy taka analize
In particular, we asked about the experience, domains, type of applications, methods, size of the project, and size of the organization. The answers to the questions could be provided using ordinal scales or as multiple-response true/false questions (more than one choice was allowed). For the questions concerning the domains and application types, we had adopted the classification scheme used by ISBSG~\cite{isbsg}, while for the size of the organization, we used the classification defined by the European Commission~\cite{euSize}. 

The last part of the questionnaire asked the respondents to leave any comments, questions, or remarks. We also asked them to voluntarily share their DoD and the problems they met in some other project caused by the lack of DoD. Optionally, every respondent might have provided their email address to get a summary of the results after the survey is completed.

\subsection{Survey instrument validation and evolution}

The prepared questionnaire underwent multiple internal and external reviews. First, it was examined by the authors of this paper and the initial version was subjected to a pilot study with 7 participants (four members of our research group and three external agile software development professionals). Their feedback allowed us to improve the wording of some questions, and a few spelling mistakes were found. One wording issue concerned that we asked about ``software projects'' and two practitioners had doubts if they are eligible to participate in the survey. They pointed out that agile approaches are also used in the context where there is no defined period of time to finish the work (projects) and they are usually called ``product teams''. Next, we found out that many software engineers have the same opinion, e.g., Sriram Narayan or Martin Fowler \cite{folwerProductProjects}. Thus, we added the information to the questionnaire. However, no major problems were identified. The questionnaire was not further modified during the study.

\subsection{Ethical considerations}

While designing our study, we have considered a series of ethical considerations, especially those discussed by Vinson and Singer \cite{vinson2008practical}.

\emph{Informed consent.} Participation in the study was voluntary. The invitation letter and the introduction page of the questionnaire form informed potential participants about the goal of the study, the research group conducting the study, the research procedure (including information on how the responses will be processed, and the results communicated), the benefits of participating, the estimated time to complete the survey, and the contact e-mail addresses of the members of the research team. 

\emph{Anonymity.} Participation in the study was anonymous. We did not ask about any personal information or the names of the companies. However, the participants could provide us with their e-mail addresses that might contain their names or surnames. Therefore, we excluded these data from further analyses.

\emph{Beneficence.} A direct benefit of participating in our study was the early access to the survey results before they are officially published. 

\emph{Confidentiality.} The online survey was conducted using the Survey Monkey platform, which we consider to be a secure service. From our analysis of the security features, it follows that they are ISO 27001 certified, EU and US Privacy Shield Certified. They use AES 256 based encryption and declare to implement appropriate technical, organizational, and administrative systems, policies, and procedures to ensure the security, integrity, and confidentiality of the questionnaire and collected data to mitigate the risk of unauthorized access to or usage. The only sensitive data concerned e-mail addressees of the respondents. They were processed separately from the remaining data and used only to send the results back to the respondents.

\subsection{Data collection}
\label{sec:data-collection}

Since there is no one `place' that provides access to the representatives of the population, we decided to target the respondents using Internet-based channels. We decided to conduct an invitation-based online survey, and, thus, we sent messages to people we knew to have experience in Agile and posted a request to participate in the survey in social network groups related to Agile on LinkedIn, Facebook, and MeetUp. After three to four weeks, we sent kind reminders to the groups.
The data collection took place between February 21, 2020 and September 21, 2020.  The exact response rate cannot be calculated due to the usage of public invitations, but given that we collected 276 responses, the response rate can be interpreted to be low, which is characteristic to online surveys.

There were 137 complete and 139 incomplete responses. The respondents spent on average ca. 17min. (median) to complete the survey (the maximum duration was 72h 31min. while the minimum duration was 4min.). There were three subgroups among those who have not completed the survey (we will refer to them as incomplete respondents). The majority of incomplete respondents 69\% (96) answered just the first question, that is about the year of their project and devoted to the survey ca. 1min. (median). Another 26\% (36) of incomplete respondents spent ca. 4min.  (median) and answered the next question---the first question that assessed the benefits of DoD. Finally, 5\% (7) of incomplete respondents dedicated ca. 7min. to the survey and left it after sharing their opinion about the problems concerning using DoD. Thus, it seems that the major issues that could discourage our respondents were either (1) the topic of the survey that after reading the introduction was not compelling enough to continue, or they did not have experience in the area, (2) the lists of benefits gave not a good impression, e.g., seemed time-consuming or complicated, or (3) answering the first three groups of questions (year, benefits, problems) made respondents not willing to continue, e.g., tired or bored.

\subsection{Data analysis methods}
\label{sec:dataAnalysis}
\begin{figure*}
\centering\includegraphics[width=0.85\textwidth]{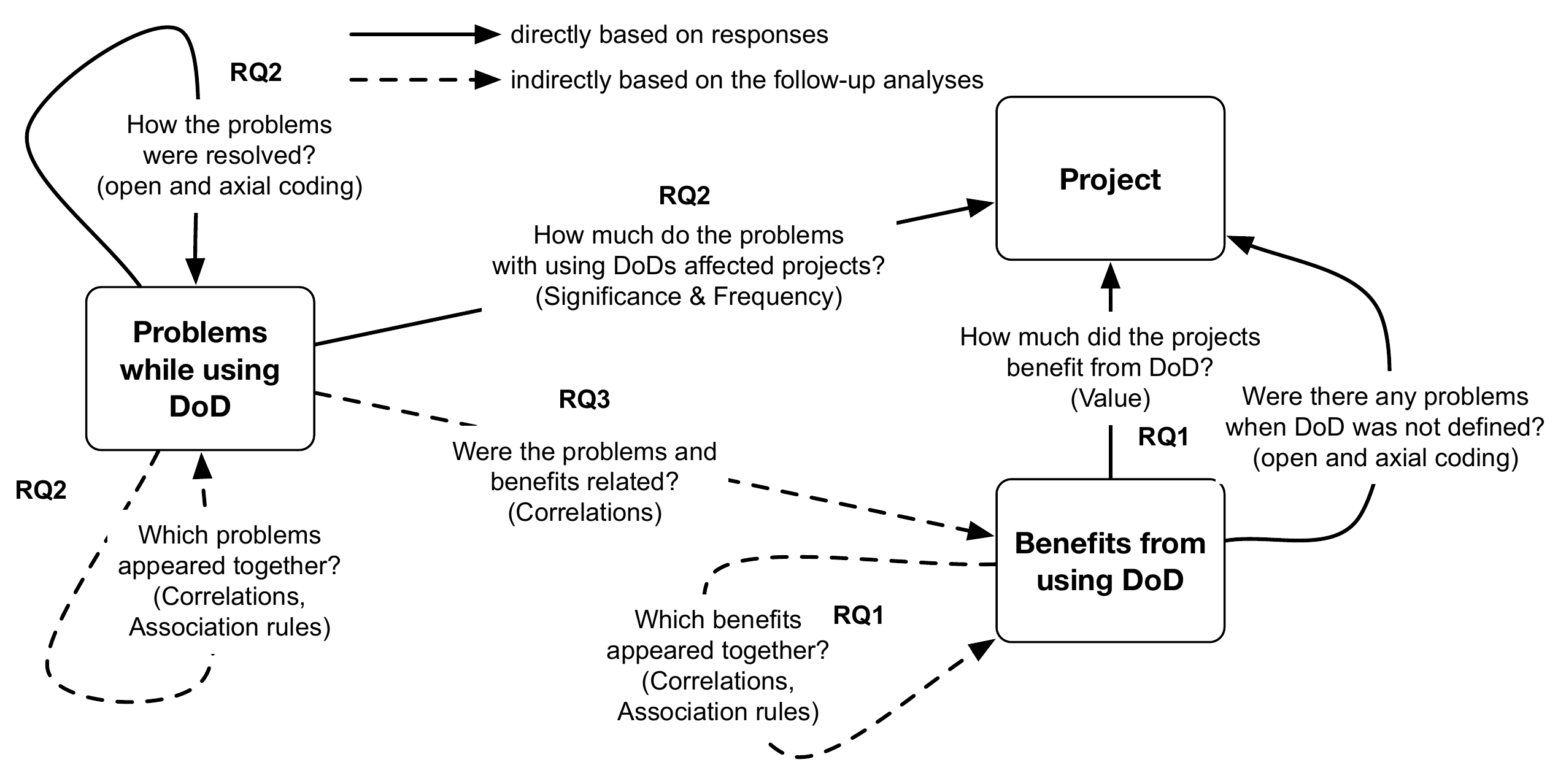}
\caption{\abc{Analyses performed and methods used to answer RQ1, RQ2, and RQ3.}}
\label{fig:analysis-dod}
\end{figure*}

\abc{We used the following criteria to validate responses: (1) a respondent has to answer all obligatory questions (to eliminate drop-outs), (2) the answers to all open questions are relevant (to eliminate misleading data), and (3) the last participation in the project the respondent focuses on in the survey was no earlier than 2015. This validation approach resulted in having no missing data and allowed us to apply quantitative analysis methods. Figure \ref{fig:analysis-dod} presents a map of analyses performed to answer RQ1--RQ3 and the analysis methods employed for that purpose.}

\abc{We used frequency analysis to analyze the direct responses to the multi-choice questions. For the questions related to problems and benefits, we introduced  additional measures by normalizing the number of responses with respect to the total number of responses (we introduce each of these measures in Section \ref{sec:results} while discussing the results). Also, we used correlation analysis and association rules to analyze the relationships between benefits (RQ1), problems (RQ2), and both of them (RQ3). In particular, we decided to base the correlation analysis on the Spearman rank-order correlation coefficient (Spearman's $\rho$) since the responses were expressed on ordinal scales. We employed statistical inference testing to filter the relationships that are unlikely to be observed by chance only (we set the significance level $\alpha$ = 0.05) and followed the guidelines for interpreting the effect size of relationships provided by Akoglu \cite{akoglu2018user}. Finally, we augmented the correlation analysis by mining association rules with the use of the Apriori algorithm \cite{agrawal1994fast} as a tool for identifying new or confirming the previously identified multi-factor relationships between the problems and benefits. When mining association rules, we converted the ordinal response scales to dichotomous scales to express whether a given benefit/problem was present or absent in a project (positive responses, e.g., ``rather significant'' and ``definitely significant'',  were mapped to ``one'' while remaining scale items were replaced with ``zero''). We set the minimum support to 0.1 (support is an indication of how frequently the itemset appears in the dataset) and confidence to 0.9 (confidence is the likelihood that item Y is present if item X is present). Unfortunately, the number of inferred rules might be too large to allow for their direct analysis. Therefore, we choose correlation analysis as our main tool for performing the analyses and consider association rules as a supporting one. Finally, although we have to emphasize that identifying a relationship (correlation) between factors does not indicate the existence of causality between them, we attempted to hypothesize about the possible reasons behind the observed relationships.} 

\abc{For the open-text questions we used the grounded theory techniques of coding (open and axial coding) and constant comparison as recommended by Charmaz \cite{charmaz2006constructing}.} %The analysis of the responses to the question that concerns providing the contents of DoD would be analyzed separately since it is not directly connected with the aim of the research study. 
\abc{
To assure reliability of the coding process, we executed the following steps. \\
\noindent Step 1: Preparation---one of the authors (the 3rd author) extracted all responses to open-text questions and moved them into separate files (one file per each question).}\\
\abc{\noindent Step 2: Individual coding--two researchers (the 1st and the 3rd author) performed (individually) open coding of the text responses, and each code summarized a single key concept. For example in the answer of respondent R15 to the question Q10 about the process of creating DoD:``\textit{\colorbox{blue!30}{Initially we had a generic template.} Then, over time \colorbox{green!30}{we discussed} each point and adjusted it to our situation. We also discussed whether there is something missing and added our own points.}'', the codes of \colorbox{green!30}{\texttt{Team}} and \colorbox{blue!30}{\texttt{Initial template}} were identified.}\\ 
\abc{\noindent Step 3: Creating the coding schema---the two researchers compared their coding schemes and discussed their findings. They applied constant comparison to group similar codes. As a result, they created the final coding schema and formulated guidelines on how to interpret and analyze the quotes. In the example from Step 2, the coding schema of categories and subcategories was developed containing:\\
{\small $\bullet$ \texttt{Who creates DoD?} \textit{\textcolor{gray!70}{(Category)}}\\
  $\rightarrow$ \texttt{Team, Product Owner, Scrum Master} \textit{\textcolor{gray!70}{(SubCateg.)}}\\
\texttt{$\bullet$ What is used to create DoD? \\
$\rightarrow$ Initial template, Standard of organization \\
$\bullet$ What activities? \\
$\rightarrow$ Adjust}\\
etc. }}\\ 
\abc{\noindent Step 4: Coding confirmation and final coding---one researcher (the 1st researcher) performed coding using the final coding scheme and the guidelines.}\\ 
\abc{\noindent Step 5: Validation of the coding schema---yet another author of the paper (the 2nd author) went through the data and the coding schema, which resulted in clarification of three codes descriptions and not finding any coding errors.}\\
\abc{\noindent Step 6: Identification of the relationships between codes---two researchers (the 1st and the 2nd author) were identifying relationships between and within codes. For example, we identified several activities concerning using DoD in all answers to the question 10 mentioned in Steps 2\&3 but also in other open text questions, such as \texttt{Adjust template}, \texttt{Discuss}, \texttt{Document}, \texttt{Analyze}. In this step number 6 we analyzed them to investigate which one follows or proceeds others, which shall be treated as obligatory, and discover any conditions under which the relationships hold. Although, the step was applied to all open questions, it turned out to be the most valuable for question Q10 concerning the process of using DoD.}\\

\abc{To analyze the responses to the open question asking to voluntarily provide the DoD, we used the the following four step approach:}\\
\noindent\abc{Step 1: Preparation---the first author of the study extracted DoD items from the answers and placed them in a table in an Excel spreadsheet.}\\
\noindent\abc{Step 2: Categorization---the first author assigned the category proposed by Silva et. al in \cite{Silva2017} and \cite{Silva2018} to each DoD item. The second author went through the assigned items and verified the correctness of assignment. After verification, during a meeting 4 changes were introduced.}\\
\noindent\abc{Step 3: Subject identification---the first author extracted subjects from each DoD item, next, the second author went through the subjects assigned to DoD items and verified the correctness of assignment. During a meeting of the authors, no change was introduced, but it was decided that one item can tackle two subjects.}\\
\noindent\abc{Step 4: Similarity and variability analysis---the first author extracted statements from same or similar DoD items and noted them down using the NoRT notation~\cite{kopczynska2018empirical}. Next, the second author went through the statements and DoD items and was to verify if he can formulate each DoD item using the assigned statement. During a meeting of the authors to discuss the results of their work no change was introduced.}\\

% jesli raportujemy w mrcv to tak
%We decided to combine several techniques to analyze and visualize the responses. 
%The most basic among them is a frequency analysis which is used to obtain a general overview concerning every question. We used statistical tests to find relationships between the demographic data, including a test for simultaneous pairwise marginal independence (SPMI) for the multiple response categorical variables (MRCV).

\subsection{Validity threats}
\label{sec:threats}
The analysis of validity threats is based on the guidelines provided by Wohlin et al.~\cite{Wohlin_etal_2012}.

\textit{Construct validity.} We identified several threats to construct validity of our study. The first one relates to the understanding of the term ``Definition of Done'' by the participants.  To mitigate this threat, we provided a definition  of DoD (based on the Scrum Guide) in the questionnaire.  It does not guarantee that all participants had a common understanding of that term; however, it should prevent the situation when a person confuses the term with some other related concepts (e.g., acceptance criteria).

Since none of the previous studies reported comprehensive lists of benefits or problems related to using DoDs, there is a threat that our list is largely incomplete or partially irrelevant. To mitigate this threat, we based the proposed list of benefits and problems on the literature regarding DoD / requirements and brainstorming sessions among the authors of this paper. We also allowed for the ``other'' answer so that the respondents could name the benefits and problems that were not on our lists. Therefore, if our list was missing some important benefits or problems, we would most likely be informed about it by at least some participants.   

As a respondent might not be aware of a certain benefit or problem, we allowed for the ``I don't know'' answer to the multichoice questions regarding benefits and problems. Also, we introduced a measure called \textit{Awareness (A)} that is calculated as the percentage of the number of answers other than ``I don't know'' with respect to the total number of answers for each question regarding benefits or problems. Observing low \textit{Awareness} would indicate that the respondents had difficulty in reliably assessing whether or not a given benefit/problem appeared in their project. This could be caused by a vague definition of the benefit/problem, respondents' lack of knowledge, or the elusive nature of the benefit/problem. 

We asked the participants if the problems they encountered had \emph{significant} negative impact on their projects,  and if using DoD was \emph{valuable} for their projects. We have to accept the fact that every person might have a different understanding of what these terms mean in this context.

Also, we made sure that the participants understood the goal of the study by clarifying the goal at the beginning of the questionnaire, so they were motivated to provide comprehensive and true answers to the questions.

Finally, we decided to run the survey anonymously, only after respondents completed all research-related questions they were asked to voluntarily provide their email address to receive a report of the results (still it gave the possibility to stay anonymous). In this way, we mitigated the evaluation apprehension threat so that the participants were guaranteed that they could be sincere about all their answers.

To mitigate the risk concerning \textit{Content validity}, i.e., the questions we asked our respondents are not representative of what they aim to measure, we asked our experts in the pilot study if they see any necessary changes to introduce to achieve the goals of our study. We need also to accept that we could have added some more questions to the questionnaire. However, there is always a trade-off between the number of questions respondents would be willing to answer and the thoroughness of answering the research questions. In our opinion and taking into consideration the exploratory goal of our study, we selected the most important questions and left the space for future research to explore, e.g., the causes of the identified problems.

\textit{Internal validity.} Although we did not seek to establish causal relationships, we believe that there are some threats that we can classify as belonging to internal validity. 

First of all, we partially relied on inviting members of agile social networks (LinkedIn, Facebook, MeetUp) to participate in our survey. As a result, it limited our control over the response collection process. Consequently, we were not able to determine neither the response rate nor who received our invitation. Also, there is a question about the trustworthiness of the respondents. However, since a high percentage of respondents (45\%) left their email addresses to be informed about the survey results, we expect that the topic was interesting to them, and they had no reason to intentionally provide false responses.

Also, informing the participants about the results of the study was the only incentive we offered for participating in the study. It could have a double-edged impact on the responses we collected. The use of monetary incentives could have increased the response rate, however, it could also harm the quality of the responses since some of the respondents might have been interested in completing the survey to be rewarded rather than  motivated by the will of sharing their opinions with the community.

The other threat concerns the skills required to fill in the questionnaire. We assumed that the respondents would not have problems in responding to an on-line survey which is created using one of the most popular survey tool (Survey Monkey) and consisting of a commonly-used type of questions. Moreover, we assumed that they are fluent-enough in English to understand the questions. We conducted a pilot study to ensure that the questionnaire is easy to understand.

We continuously monitored the process of filling in the survey (using a quick analysis of the answers in the survey tool) and, especially, the time that the respondents spent on answering the questions. We did not observe any disturbing cases and it took ca. 17min. 4sec. (median) to complete the survey, which seems to be a reasonable duration for this kind of survey (we informed the participants on the first page of the questionnaire that the estimated time of completing the survey is between 15 and 20 minutes).
We also monitored those who dropped out. Since a significant proportion of respondents left the survey after the first question (see the analysis in Section \ref{sec:data-collection}) it seems that either the topic was not interesting or they did not have experience in it, or the list of benefits of using DoD discouraged the respondents from further work (e.g., seemed time-consuming or difficult). Thus, we might suspect that those who provided complete answers were those most interested in the topic.

\textit{External validity.} The main threat to external validity concerns the representatives of the respondents and their projects. As it follows from our study design (the goal to draw an overall view of how DoDs are used) and the analysis of the demographic data (see Section~\ref{sec:results}) the respondents represent diverse profiles of software project participants (i.e., they have different experience, work in various industry sectors, projects were developed in different countries, etc.), which is essential to mitigate the risk of skewing the observations towards some particular context. The sample seems appropriate for the goal of our study, which was to get a general overview of the potential benefits and problems related to using DoDs in agile projects. However, a side effect of surveying such a broad population is that we were not able to relate certain benefits or problems to the presence of specific context factors in the projects. Therefore, based on our results, we cannot tell which benefits / problems one should expect to see in their particular agile project. 

Also, we  focused only on recent projects (i.e., the last time a respondent participated in the project was no earlier than 2015), which to some degree narrowed down the population under study and might introduce some selection bias to our study. However, our focus was not to study how the implementation of the DoD practice evolved over the years, but rather to study the current trends of how it is used and how it impacts agile software development projects.

\textit{Conclusion validity.} The threat concerns the scales used to evaluate the benefits and problems of using DoD. They are subjective and could be interpreted differently by respondents depending on their knowledge, experience, character, etc. (this is also a threat to construct validity). We also allowed for providing open-text answers or stating ``I don't know'' to avoid biasing the results by forcing the respondents to answer about the provided sets. 
We employed a qualitative coding technique to analyze open-text responses. Although, three authors of the paper performed a multi-step process of analyzing the responses (see Section \ref{sec:dataAnalysis} for details), such an approach might have introduced some bias to the conclusions.

\section{\abc{Demographic information}}
\label{sec:Demography}

As it can be seen in Table \ref{tab:demo-respondents} a), the survey respondents performed a large variety of project roles. However, most of them were responsible for Scrum Master-related and management-related tasks, programming, or requirements elicitation and analysis. Also, more than 66\% of the participants had 5 or more years of experience, while only 3\% of them worked in IT for less than a year. 

The respondents mainly referred to their recent projects since 96\% of them were developed within the last three years (69\% in 2020, 24\% in 2019, and 4\% in 2018). 

Figure \ref{fig:map} presents the countries in which the respondents' projects were developed. The most dominating areas were North America, Europe, India, and Australia, while the two most underrepresented regions were China and Central Africa. A similar geographic distribution of responses was reported by other recent surveys regarding agile methods \cite{versionone2021, ochodek2018perceived}. However, we see that, in our case, Central Europe could be overrepresented in the Europe region. 

As it is presented in Table \ref{tab:demo}, the projects were developed for different business domains, with banking and finance being the two dominating domains (Table \ref{tab:demo} a). In addition, the two most frequently developed types of applications were financial systems and web applications (Table \ref{tab:demo} b). However, we can see that the responses quite evenly covered most of the application types. The distribution of domains we observed in the studied sample is similar to the one reported by Hill \cite{hill2011practical} for the ISBSG database.

By looking at the project teams, we can see that the vast majority of them worked in Scrum (89\%), confirming that the DoD practice is strongly related to this agile framework (Table \ref{tab:demo} c). In other surveys regarding agile methods, Scrum was reported to be used by 67\% \cite{versionone2021, Matharu2015} to 94\% \cite{ochodek2018perceived} of the participants.

In nearly 60\% of the projects, there were more than ten people involved \abc{(Table \ref{tab:demo} d)}. Therefore, we can suspect that many of the projects were developed in multiteam environments. 

The analysis of the collected demographic information did not reveal strong evidence against the sample representativeness of the target population.  However, we can see that our sample of projects developed in Europe could be slightly skewed towards Central Europe.    

%For instance, multiple studies reported that Scrum is currently a dominating Agile method (e.g., \cite{ochodek2018perceived, Matharu2015, hoda2018rise}). Therefore, we  would expect to see this trend reflected in the studied sample. 

\begin{figure*}[!ht]
\centerline{\includegraphics[width=1\textwidth]{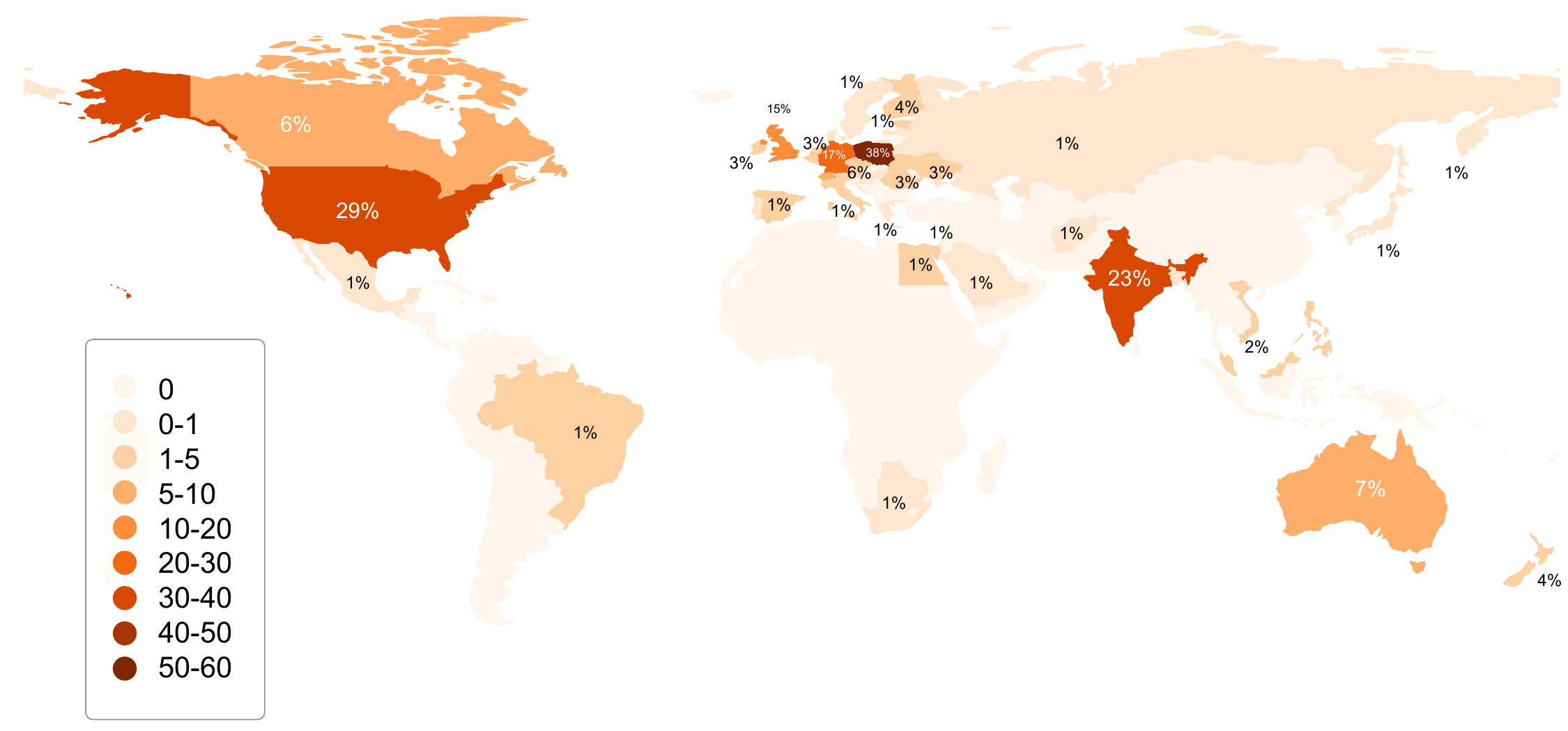}}
\caption{\abc{The countries in which the respondents' projects were developed.}}
\label{fig:map}
\end{figure*}

\begin{table}[!t]
\caption{Respondents' characteristics}
{\footnotesize
	\renewcommand{\arraystretch}{1.1}
{	\setlength{\tabcolsep}{0.35em}
	\begin{subtable}{\columnwidth}\centering{
		\begin{tabular}{p{3.1cm} | c | p{3.1cm} | c |}
			\toprule
			\multicolumn{4}{c}{\textbf{Responsibilities}} \\
			\multicolumn{4}{c}{(N=137, multiple choices allowed)} \\
			\specialrule{.4pt}{0pt}{1pt}
			Scrum Master's tasks & 68 & Designing software & 31\\
			Programming & 60 & Testing/QA & 25 \\
			Project management & 56 &  Other & 22 \\
			Requirements & 40 & &  \\
		\end{tabular}}
		\caption{\footnotesize{Responsibilities of the respondents in projects}}
	\end{subtable}
\vspace{0.25cm}

\begin{subtable}{\columnwidth}\centering{
		\begin{tabular}{p{3.1cm} | c | p{3.1cm} | c |}
			\toprule
			\multicolumn{4}{c}{\textbf{Experience}} \\
			\multicolumn{4}{c}{(N=137, single choice allowed)} \\
			\specialrule{.4pt}{0pt}{1pt}
			0--1 year & 4 & 5--10 years & 14\\
			1--3 years & 27 & over 10 years & 77 \\
			3--5 years & 15 &   &  \\
		\end{tabular}}
		\caption{\footnotesize{Experience of the respondents in projects}}
	\end{subtable}
\vspace{-0.5cm}

\label{tab:demo-respondents}
}}
\end{table}

\begin{table}[!ht]
\caption{Projects' characteristics}
{\footnotesize
	\renewcommand{\arraystretch}{1.1}
{	\setlength{\tabcolsep}{0.35em}
	\begin{subtable}{\columnwidth}\centering{
		\begin{tabular}{p{3.1cm} | c | p{3.1cm} | c |}
			\toprule
			\multicolumn{4}{c}{\textbf{Domains}} \\
			\multicolumn{4}{c}{(N=137, multiple choice allowed)} \\
			\specialrule{.4pt}{0pt}{1pt}
			 Banking & 33 &  Government  & 5 \\
			 Financial & 16 &  Insurance  & 4\\
			 Services & 13 &  Trading  & 3 \\
			 Telecommunications & 12 &  Construction   & 2  \\
			 Medical \& health care& 12 &  Entertainment  & 1 \\
			 Manufacturing & 9  &  Others  & 20 \\
			 Electronics \& computers & 7 &    & \\
			 
		\end{tabular}}
	\caption{\footnotesize{Domains in which the respondents' projects were conducted}}
	\end{subtable}
\vspace{0.25cm}	

	\begin{subtable}{\columnwidth}\centering{
		\begin{tabular}{p{3.1cm} | c | p{3.1cm} | c |}
			\toprule
			\multicolumn{4}{c}{\textbf{Types of applications}} \\
			\multicolumn{4}{c}{(N=137, multiple choice allowed)} \\
			\specialrule{.4pt}{0pt}{1pt}
			
			Financial & 18 & Sales and marketing & 10 \\
			Web or e-business & 18 & Logistics & 4 \\
			Document management & 16 & Trading & 3 \\
			Management information (MIS) & 13 & Mobile application & 1 \\
			Transaction or production  & 13 & Personnel & 0 \\
			Electronic data interchange & 12 & Other & 0\\
			Billing & 11 &  & \\

		\end{tabular}}
	\caption{\footnotesize{Types of applications developed in the respondents' projects}}
	\end{subtable}
\vspace{0.25cm}
	\begin{subtable}{\columnwidth}\centering{
		\begin{tabular}{p{3.1cm} | c | p{3.1cm} | c |}
			\toprule
			\multicolumn{4}{c}{\textbf{Methodologies}} \\
			\multicolumn{4}{c}{(N=137, multiple choices allowed)} \\
			\specialrule{.4pt}{0pt}{1pt}
			Scrum & 122 & DSDM & 1\\
			Kanban & 8 & XP & 2  \\
			Other & 4 & Crystal Clear Methods & 0 \\
		\end{tabular}}
		\caption{\footnotesize{Methodologies used in the respondents'  projects}}
	\end{subtable}
\vspace{0.25cm}
	\begin{subtable}{\columnwidth}\centering{
			\begin{tabular}{p{3.1cm} | c | p{3.1cm} | c |}
				\toprule
				\multicolumn{4}{c}{\textbf{People participating in projects}} \\
				\multicolumn{4}{c}{(N=137, single choice allowed)} \\
			\specialrule{.4pt}{0pt}{1pt}
				up to 3 people & 4 & 19--27  people & 21\\
				3--9  people & 52 & over 27  people & 27 \\
				10--18  people & 33 & & \\
		\end{tabular}}
		\caption{\footnotesize{Sizes of teams in the respondents'  projects}}
	\end{subtable}
% \vspace{0.25cm}
% 	\begin{subtable}{\columnwidth}\centering{
% 			\begin{tabular}{p{3.65cm} | c | p{3.65cm} | c |}
% 				\toprule
% 				\multicolumn{4}{c}{\textbf{Organization size}} \\
% 				\multicolumn{4}{c}{(N=137, single choice allowed)} \\
% 			\specialrule{.4pt}{0pt}{1pt}
% 				micro (up to 10 employees) & 7 & medium (50--250 employees) & 37\\
% 				small (10--50 employees) & 15 & large (250+ employees) & 78 \\
% 		\end{tabular}}
% 		\caption{\footnotesize{Organization size of the respondents' projects}}
% 	\end{subtable}
\vspace{-1cm}

\label{tab:demo}
}}
\end{table}

\section{\abc{Results and discussion}}
 \label{sec:results}
 
\subsection{Process of creating and maintaining DoDs}
\label{sec:createMaintainDoD}

\begin{figure*}[!ht]
\centerline{\includegraphics[width=1\textwidth]{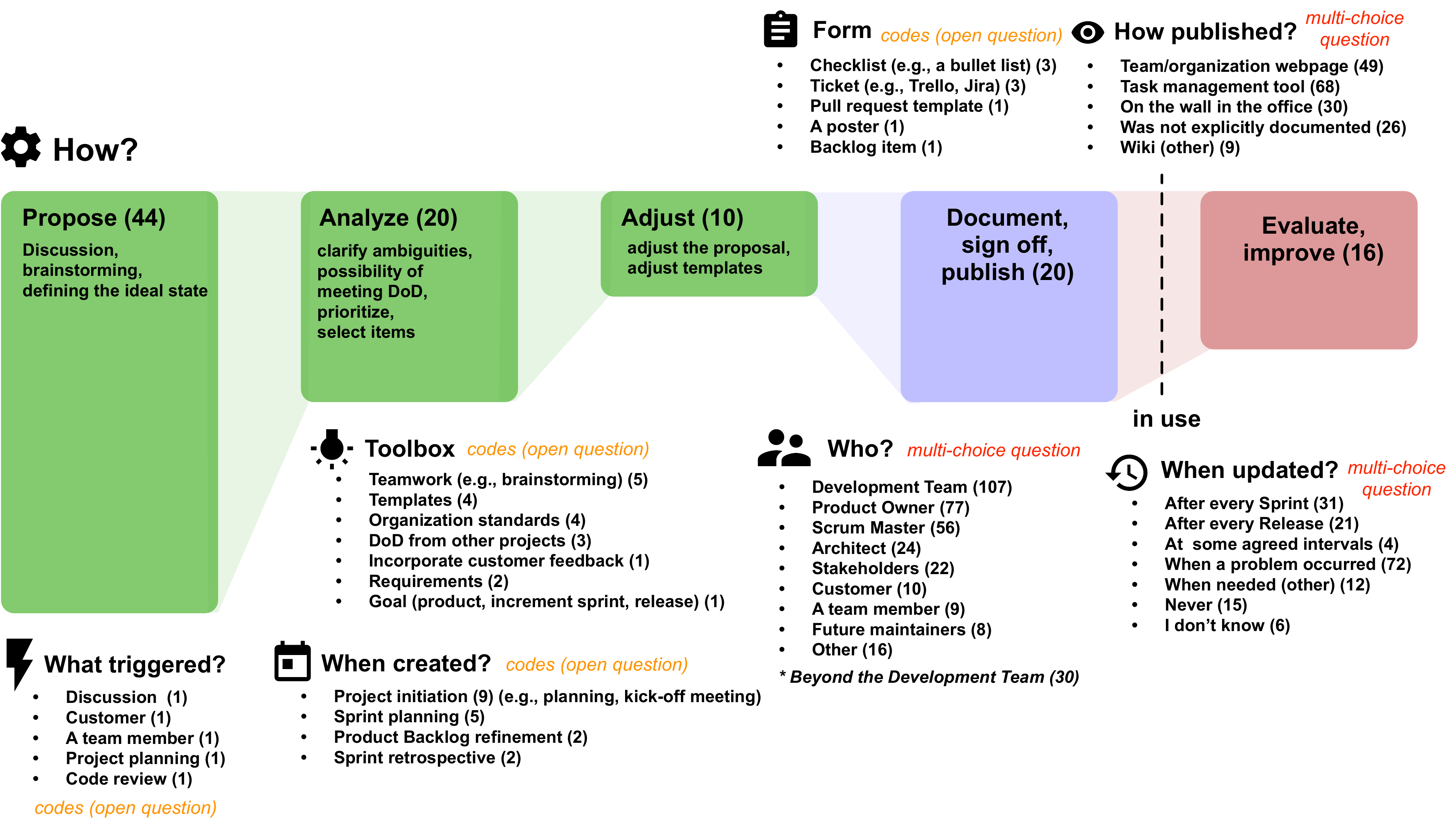}}
\caption{Process of creating and maintaining DoDs (codes from the axial coding and answers to multi-choice questions).}
\label{fig:how}
\end{figure*}

The results of the analysis of the open questions (regarding the process of creating and maintaining DoDs) and three closed questions (concerning the roles involved in that process, the frequency of updating DoDs, and the ways of publishing them) are summarized in Figure \ref{fig:how}. Also, the figure reports the number of answers to the multichoice questions and the frequency of codes resulted from the analyses of the responses to the open questions. However, since responding to the open questions was optional, and we cannot guarantee the completeness of each provided answer (i.e., mentioning all relevant concepts in the considered project), we will not attempt to evaluate the importance or popularity of the mentioned concepts based on the frequency of the codes. Still, the frequency analysis is possible for multichoice questions.

Based on the analysis of the codes regarding the process of creating DoDs, we identified that the activities mentioned by the participants create a process consisting of three stages (I--III) and five steps (1--5) (see Figure \ref{fig:how} Part How): I. (1) proposing a DoD, (2) analyzing DoD items, (3) adjusting DoD items, II. (4) documenting, agreeing on, and publishing the DoD, and III. (5) evaluating and improving the DoD. 

\abc{
Forty-four respondents stated that the process of creating DoD in their projects started from preparing a DoD proposal, e.g., by a single team member ``\textit{Single member post item in backlog with DoD}'', or by the whole team ``\textit{Initially look at what Utopia would look like}''. Also, a few of them shared the triggers that initiated the process. The process of creating DoDs could be \textbf{triggered} by project events like project planning or code reviews,  discussion between team members, or be a result of a single person's initiative (either a customer or team member). According to the respondents, the work on a DoD is equally distributed between the early stages of product development (e.g., project planning or kick-off meetings) and Sprints (e.g., Sprint planning, retrospective, and backlog refinement meetings).  The most often used \textbf{toolbox (means)} of supporting this process are teamwork techniques like brainstorming (e.g., ``\textit{We sit together and defined it}'', ``\textit{Brainstormed possible items}) or reusing knowledge from previous projects in the form of templates, organization standards, or DoD items from the past projects (e.g.``\textit{I copy-paste DoD from other projects as good starting point}''). Some DoD items are also defined based on requirements (e.g., based on non-functional requirements). 
}

\abc{
The second step of the DoD process (marked also with green in Figure~\ref{fig:how}), according to 20 participants, is a further analysis after the initial proposal is created. The goal is to clarify ambiguities, assess the feasibility of the DoD items, prioritize them, and, finally, select those that should be preserved (e.g., ``\textit{then discussion with Team where still adjustments can happen}'', ``\textit{then voted on must-haves vs. nice to haves}'', ``\textit{then a dedicated ``refinement sessio'' on DoD is organized (the sooner the better) in which shortcomings/limitations/inflexibilities and other faults are discussed and resolved}''). After analyzing the proposed DoD, teams introduce the necessary changes. Once the DoD document is ready, it is made available in the second stage of the DoD process. According to the respondents, it usually has a \textbf{form} of a checklist that is \textbf{published} on the team's web page or wiki. Another form is to store it as ``tickets'' in task management tools or as backlog items. Some teams prefer to have DoDs printed out and hang on the wall. Also, some parts of DoD could be expressed in other forms---e.g., as pull-request templates. Still, 26 respondents stated that in their projects DoDs were not explicitly documented. 
}

\abc{
Finally, the last step of the DoD process (see Figure~\ref{fig:how} the last step colored with red), the respondents mentioned was about the need for further maintenance and improvement of DoDs in their projects. In most cases, the DoDs were updated on demand whenever there was a reason to do so (e.g., ``\textit{then retrospected at various points to make tweaks}''). However, performing regular DoDs reviews (e.g., after every Sprint or release) was also a frequently reported practice (e.g., ``We refine/revisit occasionally in team retros''). Interestingly, 15 respondents stated that the DoDs in their projects have never been updated.
}

\abc{
According to the respondents, the three roles defined in Scrum (Developers, Product Owner, and Scrum Master) are most often involved in the process of creating DoDs (multichoice question, see Figure~\ref{fig:how} Who?). However, we could still see that in 22\% of the projects, the DoDs were created without involving the developers. It is a surprising observation since the DoD is used by developers everyday, and every increment they produce needs to adhere to it. 
}

\subsection{What DoD items are}
\abc{24 respondents shared with us their DoDs, which in total had 143 DoD items. To better understand what DoD items are, we analyzed them on three levels: (1) categories of DoD items, (2) what DoD items are about, and (3) contents of DoD items.
}

\abc{
First, we divided the DoD items into groups using the categories proposed by Silva et. al. in \cite{Silva2017} and \cite{Silva2018}. We noticed that we had to extend the proposal by adding one more category, namely \textit{AC check}, which would be used to group the items that concern the fulfillment of acceptance criteria.
}

\begin{figure*}[!ht]
\centerline{\includegraphics[width=0.9\textwidth]{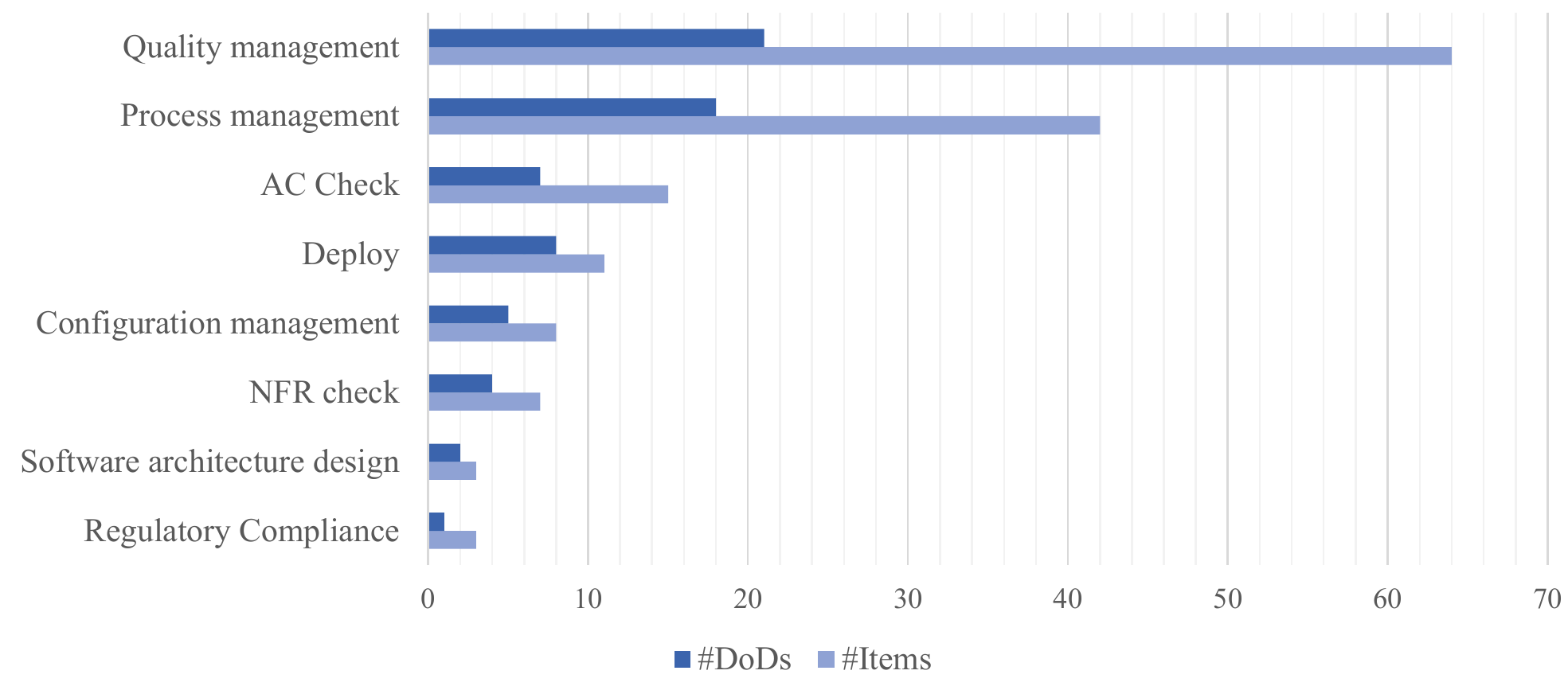}}
\caption{\abc{Number of DoD items and DoDs per category of DoD item.}}
\label{fig:categoriesDoDItems}
\end{figure*}

\abc{
As it follows from Figure~\ref{fig:categoriesDoDItems}, the greatest number of DoD items concerned Quality Management (64) and appeared in almost all DODs (21 out of 24). The second most popular category was Process Management with 42 DoD items from 18 different DoDs. The category proposed by us, AC Check, was present in 7 DoDs in 15 DoD items. There were also classified over ten DoD items (11 to be precise) from 8 different DoDs to the Deploy category. However, only one DoD contained DoD items (3) concerning Regulatory Compliance.
}

\abc{
Second, we decided to understand better what the DoD items are and extracted from each item its \textit{subject} that is the thing about which is the DoD item. It follows from Table~\ref{tab:Subjects} that DoD items are most frequently about tests (29 items), code review (15), acceptance criteria (15), and documentation (12).
}
\begin{table}[!ht]
\caption{\abc{Categories and subjects of DoD items with the number of items and number of DoDs in which they were present (multiple labels allowed).}}

{\footnotesize
	\renewcommand{\arraystretch}{1.1}
    {\setlength{\tabcolsep}{0.35em}
	\centering{
		\begin{tabular}{p{0.25cm} p{2.75cm} |c | c}
		\toprule
		 & \textbf{Subject of Item} & \textbf{$\#$ DoD Items} & \textbf{$\#$ DoDs} \\
		 \toprule
		 \multicolumn{4}{p{\columnwidth}}{\cellcolor{gray!30}{Quality Management}} \\
		 & Tests & 29& 16\\
		 & Code review  & 15 & 13 \\
		 & Code & 4 & 4 \\
		 & Feature & 4 & 3\\
 		 & Defect & 4 & 3\\
		 & (Key) Metrics & 2& 2\\
		 & Test coverage & 2 & 2\\
		 & Increment & 1 & 1\\
		 & Monitoring &1 & 1 \\
		 & Review & 1  & 1\\
		 & Technical Debt & 1 & 1\\
		 
		 \midrule
		 \multicolumn{4}{p{\columnwidth}}{\cellcolor{gray!30}{Process Management}} \\
		 & Documentation & 12 & 8\\
		 & Signoff & 7 & 5\\
		 & Feature & 4 & 4 \\
		 & Tasks & 4 & 4 \\
		 & Demo & 4& 3\\
         & Code & 3 & 3 \\
		 & Release & 2 & 2 \\
		 & User Stories & 2 & 2 \\
		 & Knowledge of team & 2 & 1 \\	 
         & Assessment & 1 & 1\\
		 & Briefing & 1 & 1 \\
	
         \midrule
         \multicolumn{4}{p{\columnwidth}}{\cellcolor{gray!30}{AC Check}} \\
          & AC & 15 & 7 \\
         \midrule
         \multicolumn{4}{p{\columnwidth}}{\cellcolor{gray!30}{Deploy}} \\
          & Deployment & 7 & 4 \\
          & Release & 3 & 3 \\
          & Release notes & 1 & 1 \\
         \midrule
         \multicolumn{4}{p{\columnwidth}}{\cellcolor{gray!30}{NFR Check}} \\
          & NFR & 6 & 3 \\
           & NFRs & 1 & 1 \\
         \midrule
         \multicolumn{4}{p{\columnwidth}}{\cellcolor{gray!30}{Configuration Management}} \\
          & Merge & 3 & 3 \\
          & Build & 2 & 2 \\
          & Code & 1 & 1 \\
          & Feature & 1 & 1 \\
          & Repository & 1 & 1 \\
            
         \midrule
         \multicolumn{4}{p{\columnwidth}}{\cellcolor{gray!30}{Regulatory Compliance}} \\
          & External standard & 3 & 1 \\
         
         \midrule
         \multicolumn{4}{p{\columnwidth}}{\cellcolor{gray!30}{Software architecture design}} \\
          & Design & 2 & 1 \\
          & API Documentation & 1 & 1 \\
        
\bottomrule
		\end{tabular}
	}
	}
}
\label{tab:Subjects}
\end{table}

\abc{
Third, we decided to better understand the contents of DoD items. We noticed that different teams (respondents) formulate either similar or the same DoD items. Thus, we focused on the similarities and variability in DoD items. To document the phenomenon we decided to use the NoRT notation, which allows to specify such differences in natural language \cite{kopczynska2018empirical}. Each statement in this notation consists of: (1) the part that is the unchangeable core; (2) parameters represented by items in angle brackets (e.g., $<$number$>$ ); (3) alternatives (options) that can be selected, are presented in brackets and are separated with bar characters (e.g., (milliseconds $|$ seconds $|$ minutes)). Each option can be either a parameter, a static (a statement that remains unchanged), or a combination of the two. To allow choosing more than one option, the ‘\}’ option modifiers are used, and to allow omitting a certain option the option modifier ']' is used.
}

\abc{
Table~\ref{tab:nort} contains 10 statements in the NoRT notation showing the similarities and variability of the DoD items. The statements are based on the DoD items most frequently mentioned by our respondents, i.e., each one is based on at least 3 DoD items (so it might be referred to as the top 10 most frequent). We added also examples of DoD items that constituted the basis for the statements, some of which were changed to make the contents anonymous.
}

\abc{
All statements but one show both similarities and variability. The highest number of similar DoD items (13) concerned the code review process. The S11 statement shows that teams either focus on a specific unit of work (e.g., task) or just generally state that the code must be reviewed. Some respondents added information about the positive result of the review process or about who will conduct the review.  The S3 statement shows that different teams use different types of testing and sometimes focus their testing effort on a defined scope (like feature) or on a certain branch. No variability was identified in the DoD items about completion of coding -- the S23 statement was formulated based on three same DoD items from different respondents.
}

\begin{table*}[!ht]
\caption{\abc{Statements showing variability and similarity in DoD items.}}

{\footnotesize
	\renewcommand{\arraystretch}{1.1}
    {\setlength{\tabcolsep}{0.35em}
	\centering{
		\begin{tabular}{p{\textwidth}}
\toprule

    $\bullet$\,\,\textbf{S11} (13 DoD items): \texttt{\{Task $|$ Code $|$ $<$unit of work$>$\} must be reviewed [and passed] [by \{at least one other person, two reviewers\}]}\\
    \;\;~e.g., \textit{Code must be reviewed by at least one other person}\\
    $\bullet$\,\,\textbf{S3} (11): \texttt{\{Unit $|$ Integration $|$ Regression $|$ Manually $|$ System $|$ $<$type of test$>$\} tested [\{feature $|$ $<$scope of test$>$\}] [on branch $<$name$>$]} \\
    \;\;~e.g., \textit{Manually tested on branch develop}\\
    $\bullet$\,\,\textbf{S14}\,\,(9):\,\,\texttt{[\{Functional $|$ Technical $|$ Integration $|$ Deployment instructions $|$  $<$name of documentation$>$\}] Docu- mentation is completed}.\\
    \;\;~e.g., \textit{Technical documentation is completed}\\
    $\bullet$\,\,\textbf{S19} (9): \texttt{The acceptance criterion is satisfied: $<$contents of the AC$>$}\\
    \;\;~e.g., \textit{The acceptance criterion is satisfied: sorting table by all columns}\\
    $\bullet$\,\,\textbf{S2} (7): \texttt{\{Functional $|$ $<$type of test$>$\} Tests were executed [and pass] [in $<$environment name$>$]}\\
    \;\;~e.g., \textit{Functional tests were executed and pass}\\
    $\bullet$\,\,\textbf{S22} (5): \texttt{Deployed to $<$environment name$>$}\\
    \;\;~e.g., \textit{Deployed to development environment}\\
    $\bullet$\,\,\textbf{S24} (4): \texttt{No \{open defects $|$ $<$type of issue$>$\} present}\\
    \;\;~e.g., \textit{No open defects present}\\
    $\bullet$\,\,\textbf{S25} (3): \texttt{Released [to $<$environment name$>$]}\\
    \;\;~e.g., \textit{Released to production}\\
    $\bullet$\,\,\textbf{S23} (3): \texttt{Coding is complete}\\
    $\bullet$\,\,\textbf{S17} (3): \texttt{$<$unit of work$>$ must satisfy the acceptance criteria}\\
    \;\;~e.g., \textit{Feature must satisfy the acceptance criteria}\\
\bottomrule
		\end{tabular}
	}
	}
}
\label{tab:nort}
\end{table*}

\subsection{Benefits of using DoDs}
\label{sec:benefits}
\begin{figure}[h!]
\centerline{\includegraphics[width=\columnwidth,trim=10 0 0 50, clip]{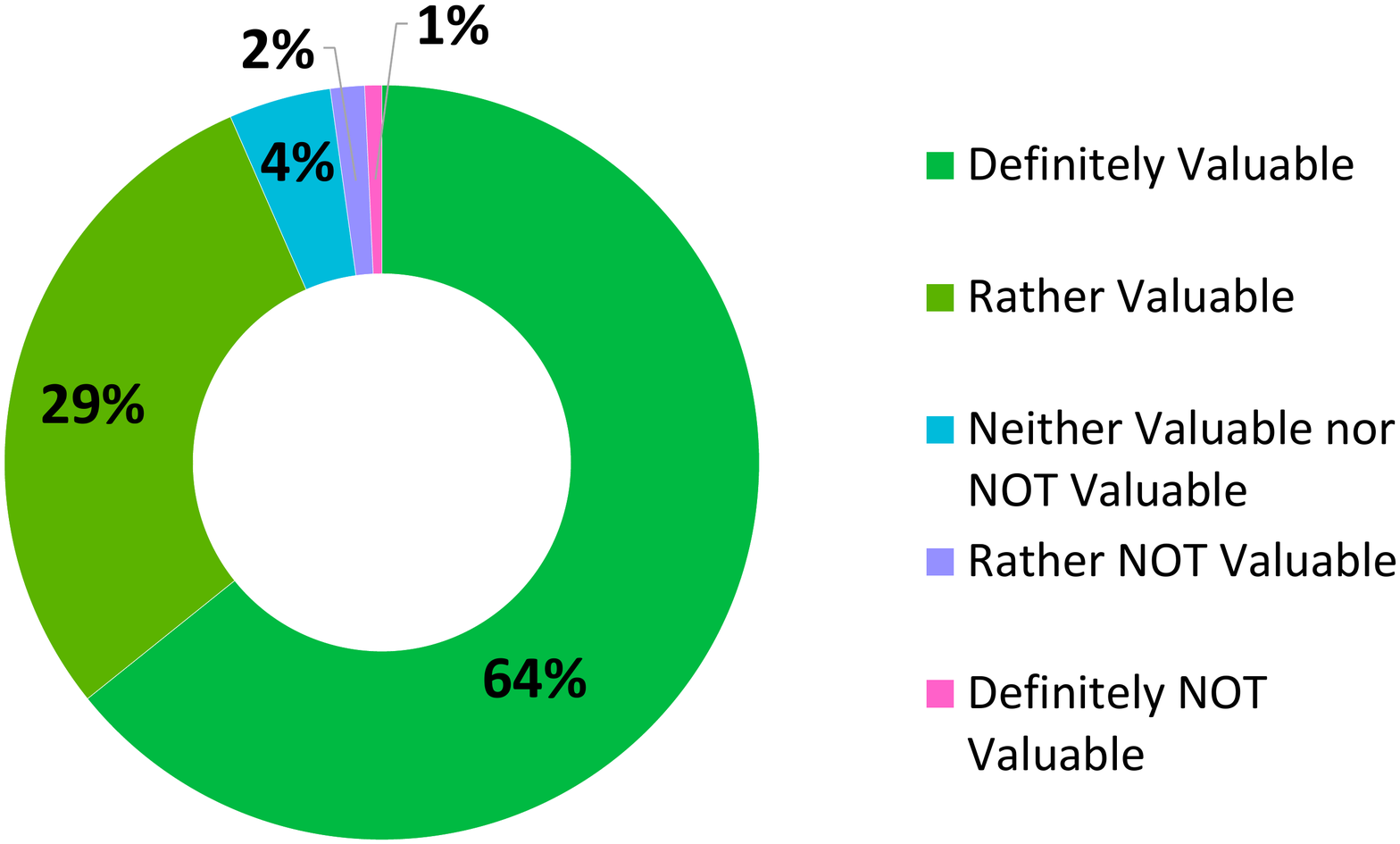}}
\caption{Perceived value of using DoD.}
\label{fig:value}
\end{figure}

As it follows from Figure~\ref{fig:value}, the great majority 93\% (128) of the respondents regarded using DoD as valuable for their project (``Rather valuable'' or ``Definitely valuable''), while 64\% of them (88) perceived it as definitely valuable.

%\begin{figure}
%\centerline{\includegraphics[width=\columnwidth,trim=40 100 50 80, clip]{figures/pievaluable.pdf}}
%\caption{Perceived valuableness of DoD.}
%\label{fig:value}
%\end{figure}

To investigate how often certain benefits of using DoDs appear in the projects, we defined a measure called \textit{Perceived Value (Val)}. It is calculated as the percentage of the answers confirming the presence of a given benefit in a project (``Definitely YES'' or ``Rather YES'') with respect to the total number of the responses, but excluding the ``I don't know'' answers. Excluding the ``I don't know'' answers from the denominator means that we focused only on the cases for which the respondents felt competent enough to judge the causality between using the DoD practice and the presence of a given benefit. Please note that there was also a neutral answer available for the respondents (``Neither significant nor NOT  significant''). Still, it does not mean that we completely ignored the ``I don't know'' answers. These answers were incorporated into the \emph{Awareness} measure which is the percentage of the number of answers other than ``I don't know'' with respect to the total number of answers for each question. It complements \textit{Perceived Value} by showing the level of support for the given benefit in the collected data. For instance, a benefit with very high \textit{Perceived Value} and very low \textit{Awareness} should be taken with a grain of salt (especially, since we search for benefits that are common across different project contexts). Of course, there might still be several valid reasons for making such an observation (e.g., a benefit might be visible only to some specific roles, or it could be intangible by its nature---be something difficult to measure).  

The top five benefits of the highest Perceived Value concern organizational (management) and technical areas are (see Table~\ref{tab:benefits}): making work items complete, assuring product quality, ensuring that the activities other than coding were executed, ensuring that all quality gates are passed, and keeping the product releasable. All but one of the benefits appeared in over 50\% of the projects. The one exceptional case was the B19 ``Less time spent on manual testing'' benefit. It appeared in around 1/3 of the projects (32\%). Finally, the respondents extended the provided list of benefits by adding two new proposals: B20 ``DoD helped with creating clean code'' and  B21 ``DoD caused that the whole team felt the ownership of quality.'' 

The calculated \textit{Awareness (A)} (see the preceding paragraphs and  Section~\ref{sec:threats} for the explanation of the measure) for the benefits ranged between 90\% and 99\%. Thus, we conclude that \emph{the respondents were well-informed and knowledgeable about the consequences of using DoD}. The benefit with the lowest \textit{Awareness} was: B9 ``Increase in customer's satisfaction.'' The second lowest \textit{Awareness} was observed for B17 ``Balance between short-term delivery of features and long-term product quality.'' Both of these benefits might be difficult to judge by a project team member who is not closely collaborating with the customer or is not involved in the planning and assessment of product quality.

\begin{table}[!ht]
\caption{Benefits of using Definition of Done}
{\footnotesize
	\renewcommand{\arraystretch}{1.1}
    {\setlength{\tabcolsep}{0.35em}
	\centering{
		\begin{tabular}{|c|p{4.6cm}|c|c|}
			\toprule
			\multicolumn{1}{|c}{\textbf{ID}}& \multicolumn{1}{|c}{\textbf{Benefit from using DoD}} & \multicolumn{1}{|c|}{\textbf{Val\%}} & \multicolumn{1}{|c|}{\textbf{A\%}}\\
			%\multicolumn{2}{c}{()} \\
			\specialrule{.4pt}{0pt}{1pt}
B1 & Help to make work items complete & 93 & 99 \\ 
B2 & Help to assure quality of product & 92 & 98 \\ 
B3 & Help to ensure that the activities other than coding were executed (e.g., code review, manual testing, build) & 90 & 99 \\ 
B4 & Help to ensure that all quality gates are passed (e.g., performance testing, code review, security check) & 90 & 98 \\ 
B5 & Help to keep product releasable & 86 & 97 \\ 
B6 & Help in effort estimation & 79 & 98 \\ 
B7 & Promotion on the meaning of ``complete'' work between stakeholders & 79 & 96 \\ 
B8 & Fewer bugs and issues get released & 76 & 98 \\ 
B9 & Increase in customer's satisfaction & 72 & 90 \\ 
B10 & Help to make team members aware of the current status of the project & 66 & 97 \\ 
B11 & Help to ensure organization's coding standards & 66 & 97 \\ 
B12 & Reduction of technical debt (early spotting the defects) & 64 & 98 \\ 
B13 & Increase of Development Team productivity & 63 & 96 \\ 
B14 & Reduction in time for reworks of implemented features & 61 & 96 \\ 
B15 & Help to keep the documentation up-to-date & 56 & 96 \\ 
B16 & Transparency of code quality (e.g., explained what it is) & 55 & 96 \\ 
B17 & Balance between short-term delivery of features and long-term product quality & 55 & 93 \\ 
B18 & Help to keep code repository clean & 52 & 98 \\ 
B19 & Less time spent on manual testing & 32 & 96 \\ 

\hline

*B20 & ``\textit{DoD helped with creating clean code}'' & -- & -- \\
*B21 & ``\textit{DoD caused that the whole team felt the ownership of quality}'' & -- & -- \\

\bottomrule

\multicolumn{4}{p{0.8\columnwidth}}{* benefits identified by some the participants (using the ``other'' option).} \\

	\end{tabular}}
	}
}
\label{tab:benefits}
\end{table}

\begin{figure*}[h!]
\centerline{\includegraphics[width=\textwidth]{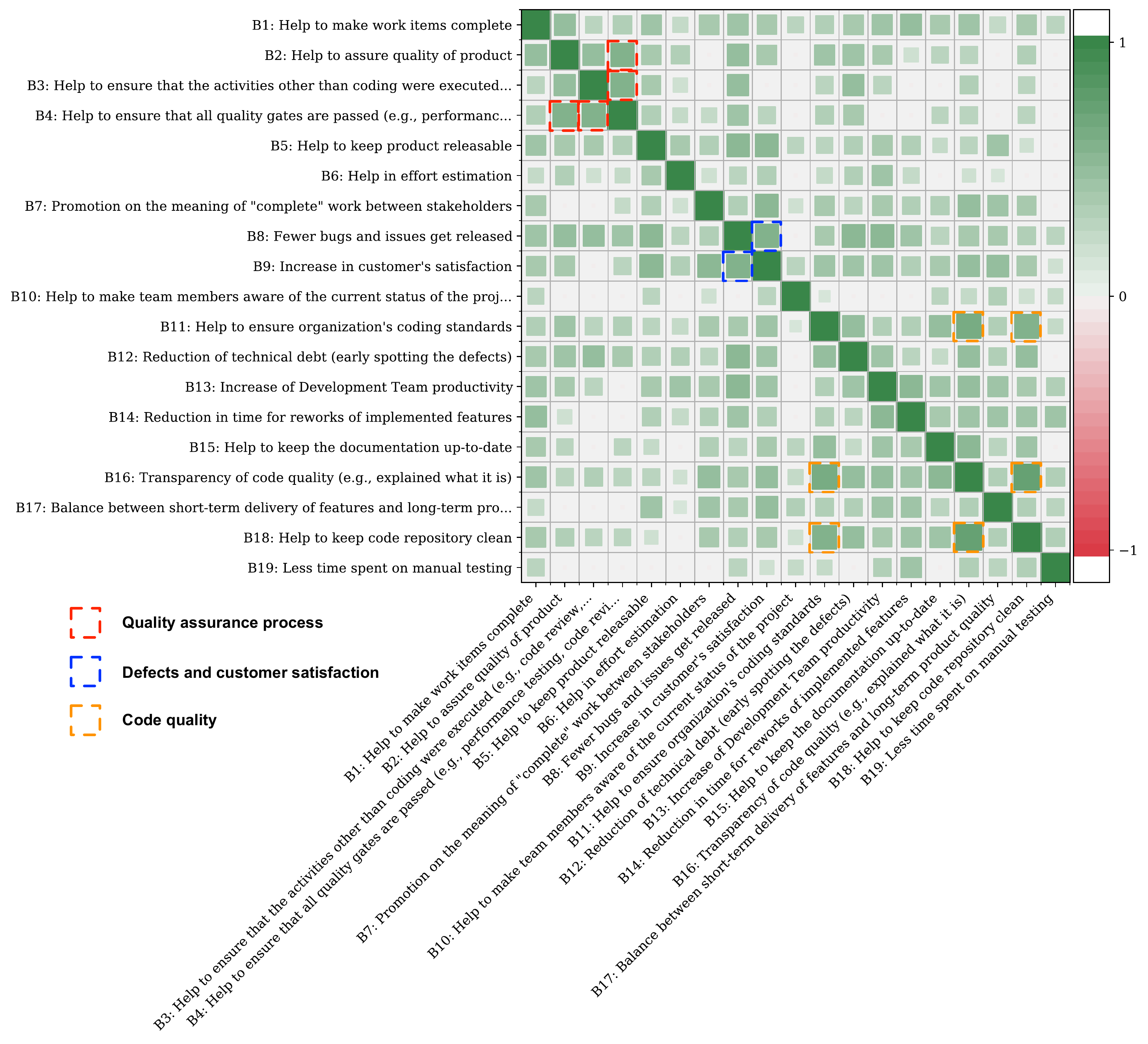}}
\caption{\abc{Correlations between the reported benefits of using DoD (only statistically significant correlations are presented).}}
\label{fig:benefits-rel}
\end{figure*}

\abc{As we can see in Figure \ref{fig:benefits-rel}, the benefits of using DoD seem to be highly correlated with each other (only positive correlations were observed). The calculated Spearman's $\rho$ for the statistically significant relationships ranged between 0.13 and 0.75 (mean = 0.39). According to Akoglu \cite{akoglu2018user} the mean value of $\rho$ = 0.39 could be interpreted as a moderate/strong relationship, while the strongest observed relationships could be classified as moderate to very strong.}

\abc{The strongest correlation ($\rho$ between 0.64 and 0.75) was observed between a group of three benefits related to code quality (B11: ensuring coding standards, B16: transparency of code quality, and B18: clean repositories). The second highly correlated group of benefits ($\rho$ ca. 0.61) regarded the quality-assurance process (B2: helping assure the quality of product, B3: ensuring that the activities other than coding are executed, and B4: ensuring that all quality gates are passed). Finally, one more correlation with $\rho \geq$ 0.60 was observed that shows a relationship between the benefit of a smaller number of defects being released (B8) and the increase in customer satisfaction (B9). Finally, there were only two benefits that did not correlate visibly with many other benefits, i.e., less time spent on manual testing (B19) and making team members aware of the current project status (B10). Unfortunately, since so many benefits were highly correlated, the number of identified association rules did not allow us to reveal any additional interesting and meaningful groups of benefits. }

\subsection{Problems resulting from not using DoDs}

When deciding on adopting a practice, it is important to know what benefits it might bring to the project, however, it is also useful to understand the consequences of neglecting it.

\begin{figure}
%\centerline{\includegraphics[width=\columnwidth,trim=5 10 350 2, clip]{figures/Issues.pdf}}
\centering\includegraphics[width=0.95\columnwidth]{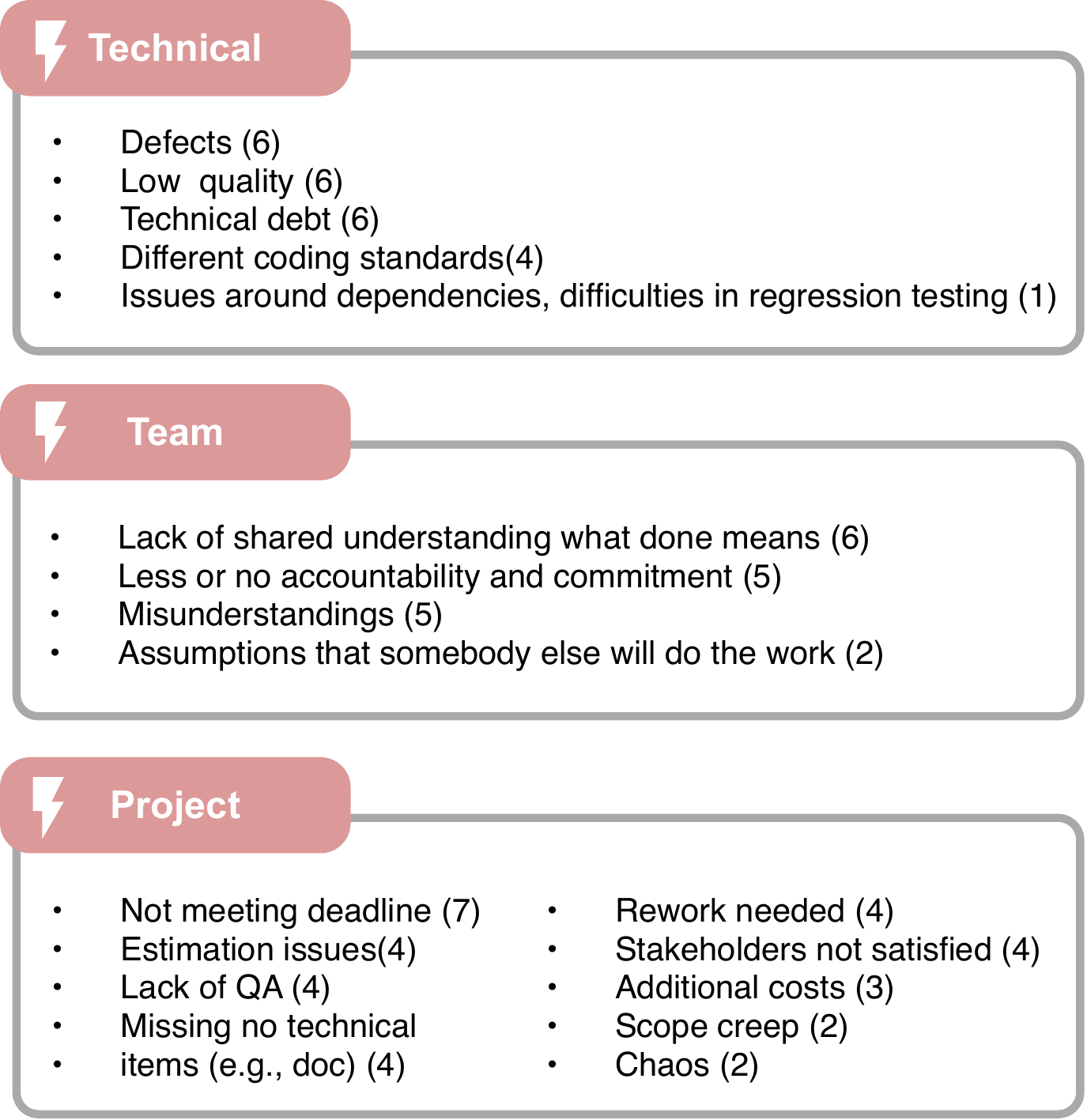}
\caption{Problems that software projects  struggle with when the DoD practice is not used.}
\label{fig:problemswhennodod}
\end{figure}

\abc{Thirty-one respondents shared some problems they had encountered in their other projects that occurred due to the lack of DoD. Those aspects concern technical-, team-, and project-level issues and are listed in Figure~\ref{fig:problemswhennodod}. The top five most frequently mentioned problems were: 
\begin{itemize}
    \item  not meeting deadlines (e.g.,``\textit{we went beyond the timeline and budget}''), 
    \item lack of shared understanding of what done means (e.g., ``\textit{Some member have different view of what to achieve, which sometimes resulting in over commit to task, team member fighting on what task to cover in each feature}'', ``\textit{Not having alignment on what ``done'' means for a team causes lots of problems and friction between dev team and PO and team and stakeholders.}''), 
    \item defects, low quality, and technical debt (e.g.,``\textit{technical debt creeping in}'', ``\textit{they haven’t delivered the expected quality and ended up in lot of rework post UAT}'', ``\textit{Finally, defects were commonplace}'').
\end{itemize}
}

\subsection{Problems encountered while using DoDs}

\begin{figure*}[!ht]
\centerline{\includegraphics[width=1\textwidth]{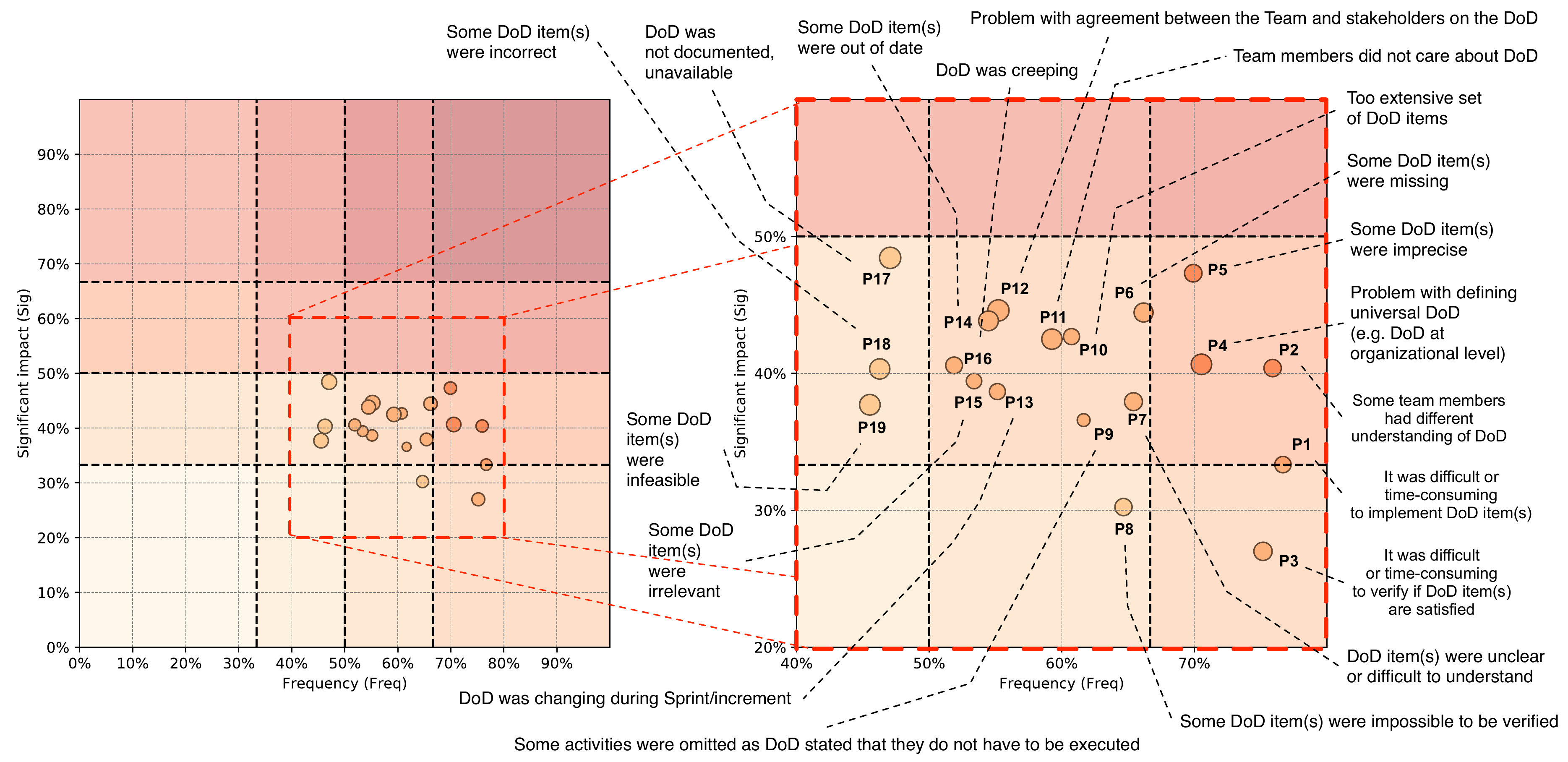}}
\caption{The importance of identified problems related to using DoDs (color---importance, size---the number of responses assessing the negative impact of problems as definitely significant).}
\label{fig:map}
\end{figure*}

\begin{table}[!ht]
\caption{Problems with using Definition of Done}
{\footnotesize
	\renewcommand{\arraystretch}{1.1}
    {\setlength{\tabcolsep}{0.35em}
	\centering{
		\begin{tabular}{| c | p{3.6cm} | c | c| c|}
			\toprule
			\textbf{ID} & \multicolumn{1}{| c |}{\textbf{Problem with using DoD}} & \textbf{Freq\%} & \textbf{Sig\%} & \textbf{A\%} \\
			%\multicolumn{2}{c}{()} \\
			\specialrule{.4pt}{0pt}{1pt}
P1 & It was difficult or time-consuming to implement DoD item(s) & 77 & 33 & 97 \\
P2 & Some team members had different understanding of DoD & 76 & 40 & 100 \\
P3 & It was difficult or time-consuming to verify if DoD item(s) are satisfied & 75 & 27 & 97 \\
P4 & Problem with defining universal DoD (e.g. DoD on organization level) & 71 & 40 & 94\\
P5 &  Some DoD item(s) were imprecise (loosely stated goals) & 70 & 47 & 97 \\
P6 & Some DoD item(s) were missing & 66 & 44 & 99 \\
P7 & DoD item(s) were unclear or difficult to understand & 65 & 38 & 97\\
P8 & Some DoD item(s) were impossible to be verified & 65 & 30 & 97 \\
P9 & Some activities were omitted as DoD stated that they do not have be executed & 62 & 37 & 97\\
P10 & Too extensive set of DoD items & 61 & 43 & 99 \\
P11 & Team members did not care about DoD (they omitted some or all DoD items) & 59 & 43 & 99\\
P12 & Problem with agreement between the Team and stakeholders on the DoD & 55 & 45 & 98\\
P13 & DoD was changing during Sprint/increment & 55 & 39 & 99\\
P14 & Some DoD item(s) were out of date & 55 & 44 & 98 \\
P15 & Some DoD item(s) were irrelevant & 53 & 39 & 97\\
P16 & DoD was creeping (continuously growing in uncontrolled manner) & 52 & 41 & 97\\
P17 & DoD was not documented, unavailable (DoD was established verbally, not written down) & 47 & 48 & 99\\
P18 & Some DoD item(s) were incorrect & 46 & 40 & 98\\
P19 & Some DoD item(s) were infeasible (impossible to satisfied) & 46 & 38 & 98\\

\bottomrule
	\end{tabular}}
	}
}
\label{tab:problems}
\end{table}

We analyzed the responses regarding the problems that might appear while using DoDs by investigating how often certain issues appeared in the projects and how harmful they were. In particular, we calculated two measures that allowed us to quantify these two aspects:
\begin{itemize}
    \item \textit{Frequency (Freq)} is the percentage of all responses except for the ``Problem did not appear'' and ``I don't know'' with respect to the total number of responses other than the ``I don't know'' answers;
    
    \item \textit{Significant impact (Sig)} is the percentage of the responses claiming that the problem had a ``Definitely Significant'' or ``Rather Significant'' negative impact on the project with respect to the total number of responses other than the ``I don't know'' answers.
\end{itemize}

\textit{Freq} tells us how likely it is that a given problem appears in a project which uses the DoD practice. We excluded the ``I don't know'' answers for the same reasons as we did it when calculating \emph{Perceived Value} (see Section \ref{sec:benefits}). These answers can either mean that the respondents had not encountered a given problem before or that they had insufficient knowledge to confirm its presence or lack of thereof. \textit{Sig} allows us to evaluate the severity of problems. We focused only on the situations when a given problem leads to serious negative consequences. However, even when the respondents stated that the impact of a given problem was not significant, it does not mean that it was negligible.

As it follows from Table~\ref{tab:problems}, the top 5 most frequent problems (P1 ``It was difficult or time-consuming to implement DoD item(s)'', P2 ``Some team members had different understanding of DoD'', P3 ``It was difficult or time-consuming to verify if DoD item(s) are satisfied'', P4 ``Problem with defining universal DoD'', and P5 ``Some DoD item(s) were imprecise'') were reported to appear in 70\% or more of the projects. The next two frequently appearing problems were missing (P6), unclear / difficult to understand (P7), or impossible to verify (P8) DoDs, which were reported to be present in over 60\% of projects. The least frequent problems, i.e., infeasible DoD items (P19), incorrect DoD items (P18), and not written down DoD (P17) were present in more than 45\% of the projects.

With respect to significance, we observed that the problems concerning the unavailability of DoD (P17), imprecise DoD items (P5), missing DoD items (P6), and reaching an agreement between the team and stakeholders (P12) were perceived as having a significant, negative impact on 44\%--48\% of the projects.

To evaluate the importance of the problems, we should consider both \textit{Freq} and \textit{Sig} measures. Figure \ref{fig:map} presents each of the problems in the two-dimensional space of these two measures with arbitrary guiding lines added for each of the measures at 33\%, 50\%, and 67\%. Based on the plot, we could state that the most important problems are P2, P4, P5, and P6 since they are all frequently appearing and have significant impact. All of them relate to the issues with specifying DoDs (missing, imprecise, or ambiguous items that are understood differently by team members). 

%Every problem that appears in every second project is definitely worth noticing, also a problem that in 1/3 of cases results in harmful consequences should not be easily neglected. We decided to slightly prefer \textit{Sig} over \textit{Freq} since the presence of even a few minor issues seems less problematic than the appearance of a single but severe problem. The resulted grouping of the problems into importance clusters is presented in Figure \ref{fig:map}. The problems that could be considered as the most important are P2, P4, and P5. All of them relate to the issues with specifying DoDs (imprecise and ambiguous items that are understood differently by team members). These problems are well-known to the requirements-engineering community. The problems related to the cost of creating DoD items and verifying that the increment meets them (P1 and P3) were also very common, however, they were less frequently causing severe problems in the projects. Less frequently appearing, but still quite significant, were the problems concerning infeasibility of DoD items (P19), their correctness (P18), and their elusiveness when DoD was not documented / made available to all team members (P17). The remaining problems were appearing in between 1/2 and 2/3 projects and were indicated as causing significant problems in 1/3 to 1/2 of the projects. 

To investigate if the findings concerning the problems are reliable, similarly like in the case of the earlier analysis of benefits, we calculated the \textit{Awareness (A)}, which was between 94\% and 100\%. Thus, it seems that the respondents were also well-informed and knowledgeable to answer the questions about the problems concerning using DoD.

%To investigate if the findings concerning the problems are reliable, similarly like in the case of the earlier analysis of benefits, we calculated the \textit{Awareness} as the percentage of the number of answers other than ``I don't know'' with respect to the total number of answers for each question asking about problems with using DoD. It turned out that \textit{Awareness} was between 94.16\% and 100\%, . Thus, it seems that the respondents were well-informed and knowledgeable about problems concerning using DoD.

\begin{figure*}[h!]
\centerline{\includegraphics[width=\textwidth]{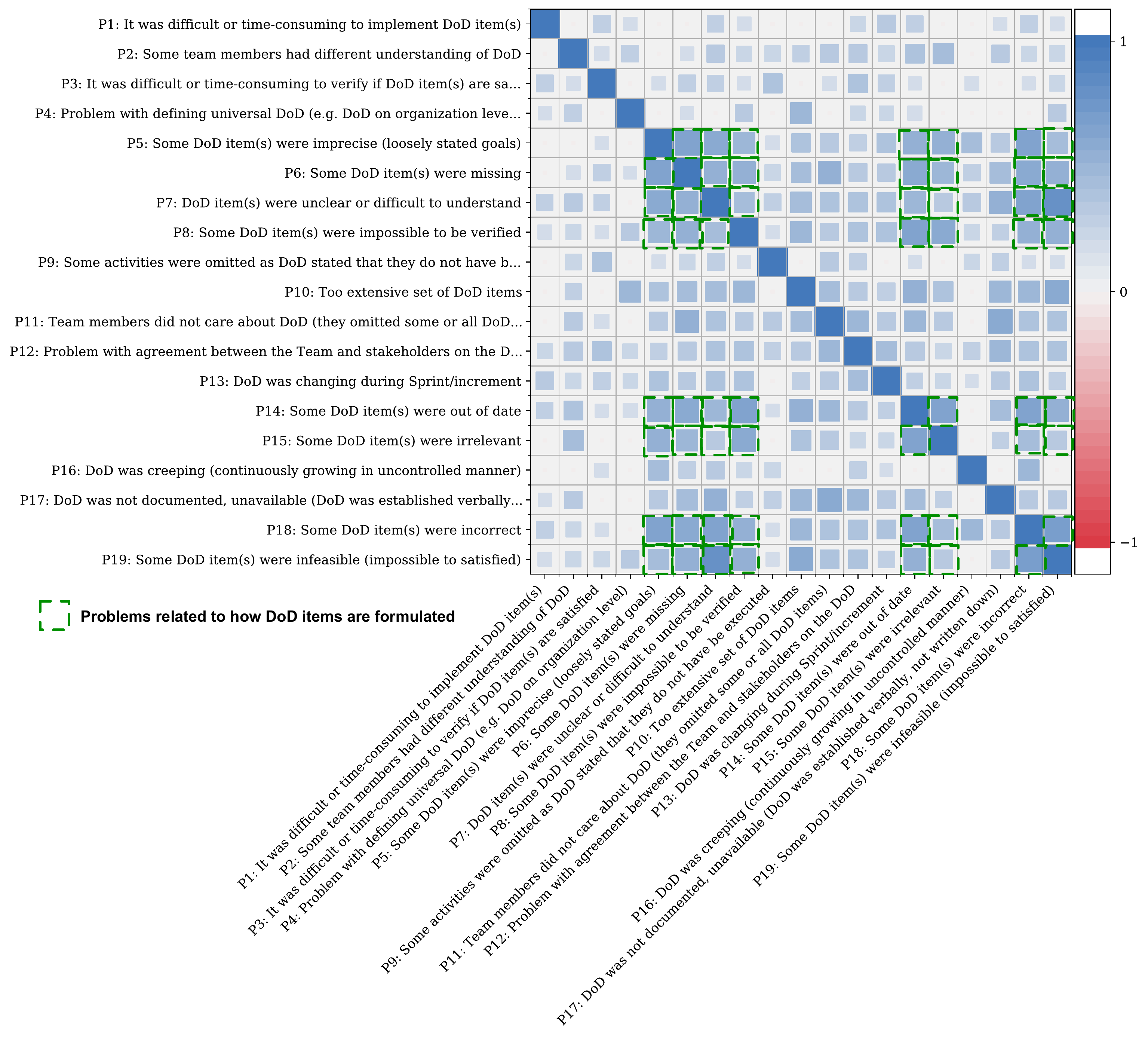}}
\caption{\abc{Correlations between the reported problems while using DoD (only statistically significant correlations are presented).}}
\label{fig:problems-rel}
\end{figure*}

\abc{Figure \ref{fig:problems-rel} shows the correlation matrix for problems while using DoD. The calculated Spearman's $\rho$ for the statistically significant relationships ranged between 0.17 and 0.80 (mean = 0.39). The mean value of $\rho$ = 0.39 could be interpreted as moderate/strong relationship while the strongest observed relationships could be classified as strong to very strong. The analysis of the frequent itemsets, association rules, and the calculated correlation coefficients allowed us to identify a group of eight problems with strong intra-group correlations. All these problems relate to how DoD-items are formulated/specified: imprecise (P5), missing (P6), unclear (P7), impossible to verify (P8), out of date (P14), irrelevant (P15), incorrect (P18), and infeasible (P19).  When we combined all these problems into a single meta-problem, it appeared to be correlated with many other problems (at least one of the problems covered by the meta-problem was correlated with another problem), e.g., a different understanding of DoD (P2), difficulties and effort while verifying DoD satisfaction (P3), too extensive DoD (P10), or problems with the agreement between team and stakeholders on DoD. We were also able to identify some other relationships by analyzing Figure \ref{fig:problems-rel} and association rules. For instance, in the association rules having ``P2: Some team members had a different understanding of DoD'' as consequent, the antecedents were (apart of the previously mentioned meta-problem with how DoD items are formulated) problems like a too extensive set of DoD items (P10), problems with agreeing on DoD within stakeholders and team (P12), difficulties and effort required to verify DoD satisfaction (P3), or problems with defining DoD at the organizational level (P4). Although these relationships are derived based on correlations, we could hypothesize that some causality relationships between these problems exist. For example, if a set of DoD items is too extensive, it might be hard to memorize it, and consequently, it could lead to a different understanding of the DoD by team members. Another example could be ``P10: Too extensive set of DoD items'' which was a consequence of rules having antecedents such as problems with defining a DoD at the level of organization (P4), omitting some activities defined in DoD (P9), problems with the common understanding of DoD (P2), and the meta-problem regarding how DoD is formulated. One more example would be ``P12: Problem with agreement between the Team and stakeholders on the DoD'' which was a consequent of rules having antecedents such as difficulties in verifying DoD items satisfaction (P3), a different understanding of DoD by team members (P2), problems with creating universal DoD (P4), changing DoD during iterations (P13), or when team members did not care about DoD (P11).}

Fortunately, many of the problems could be resolved. Twenty-three respondents shared their experience on how the problems concerning DoD were solved in their projects. The majority of them (11) explicitly stated that the problems were solved by team (e.g., ``\textit{Fix them! It's just a matter of sitting down with the team}'', ``\textit{we called in a team meeting and worked out issues}.'' Two respondents reckoned that a person with a coach/Scrum master role was needed to support the team or initiate the process.
Six respondents stated that the problems were solved while team members were interacting, mostly while team meetings/workshops such as retrospectives. What is interesting is that two respondents highlighted that solving problems with DoDs is not a ``one-shot'' activity, but an incremental process since DoDs evolve in time.
During the process of solving the problems, DoD items were clarified (5), ``\textit{started to have shared understanding}" (2), made relevant (2), adjusted (1), discussed (1), negotiated (1), made useful (1), some external dependencies were removed (1), or at one extreme the DoD was removed (1).

\subsection{Problems and benefits relationships}

\abc{The survey questions asked directly about the problems that appeared in the participants' project while using DoD and about their significance. However, by analyzing the relationships between the problems and benefits, we can investigate whether the benefits from using DoD are less/more likely to be observed when certain problems with DoD materialize (or alternatively, whether they are independent).}

\begin{figure*}[h!]
\centerline{\includegraphics[width=\textwidth]{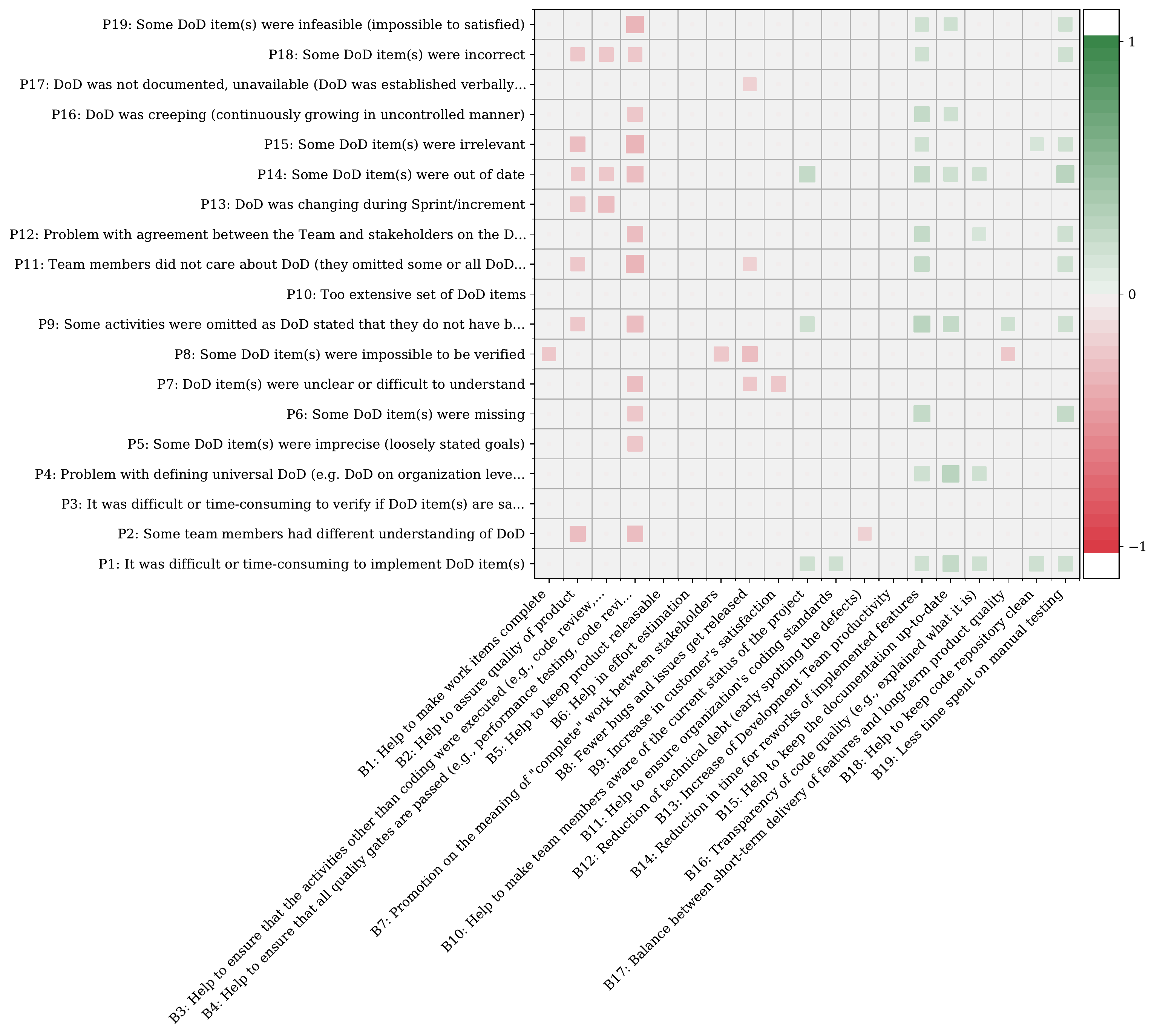}}
\caption{\abc{Correlations between the reported problems and benefits of using DoD (only statistically significant correlations are presented).}}
\label{fig:problems-benefit-rel}
\end{figure*}

\abc{The correlation matrix in Figure \ref{fig:problems-benefit-rel} shows that both positive and negative (statistically significant) correlations could be observed between problems and benefits. The calculated Spearman's $\rho$ ranged between 0.18 and 0.32 (mean = 0.23) for positive correlations and between -0.17 and -0.33 (mean = -0.23) for negative ones and could be interpreted as weak to moderate relationships.}

\abc{Observing negative correlations means that the more significant problems with using DoD are observed, the fewer benefits one should see in a project. Most of the benefits are correlated with at most a few problems. Benefit ``B4: Help to ensure that all quality gates are passed'' is an exception since it is negatively correlated with nearly all the problems. Another benefit that is negatively correlated with numerous problems (seven) is ``B2: Help to assure quality of product.'' Both of these benefits are related to the quality assurance process and both correlate with the problem ``P11: Team members did not care about DoD (they omitted some or all DoD items).'' We hypothesize that the fact that a team does not care about DoD might be an indicator that the team could be also reluctant to perform other pro-quality activities. Both of these benefits are also correlated with some problems related to how DoD is formulated (our meta-problem). Finally, we could see that benefit ``B8: Fewer bugs and issues get released'' is negatively correlated with problems such as ``P7: DoD item(s) were unclear or difficult to understand'', ``P8: Some DoD item(s) were impossible to be verified'', ``P17: DoD was not documented, unavailable (DoD was established verbally, not written down)'', and again with ``P11: Team members did not care about DoD (they omitted some or all DoD items).'' We hypothesize that there could be some causality between the problems and benefit B8 since items that are unclear, difficult to understand (P7), impossible to verify (B8), and agreed only verbally (P17) could make the testing process difficult (not to mention that the presence of problem P11 characterizes the team's attitude to quality assurance).   }

\abc{The presence of positive correlations between problems and benefits might seem surprising at first, however, we believe that some of these relationships could be justified. For instance, observing positive correlations between ``P1: It was difficult or time-consuming to implement DoD item(s)'' and numerous benefits show that ``quality does not come for free'' and teams need to invest time and resources into pro-quality activities to see the benefits. Another reason for observing positive correlations could be that the appearance of some DoD-related problems might be a good indicator of deeper problems in projects or organizations (e.g., problems with communication, decision making, or resigning from pro-quality activities)---DoD could be perceived as even more valuable by teams that struggle with quality assurance. Finally, the problems with using DoD could be temporal, and in some cases, their presence could even trigger positive changes and lead to implementing corrective actions. }

\section{\abc{Study implications}}
\label{sec:discussion}

\subsection{Implications to research}

Our study confirms that using DoDs is useful, as has already been claimed by several researchers and practitioners based on their experience, e.g., \cite{Saddington2012}, \cite{jakobsen2008mature}, \cite{cohn2010succeeding}. 

We identified and evaluated the importance of several benefits that DoDs might bring. However, our findings are a trigger for opening a new discussion about when such benefits are present and how to create an effective DoD to help materialize these benefits.

Our study extends the body of knowledge related to the problems encountered while using DoDs by providing an initial evaluation of their frequency and significance. Since several problems might appear frequently and have a negative impact on agile projects, a further and more in-depth analysis would be valuable. We identify an opportunity for more fine-grained research on why the problems appear, what we can do to prevent them, and how to solve them.
 
%As we found in our literature review, there are a few studies that report problems concerning DoDs, e.g., \cite{OConnor2010}, \cite{Igaki2014}.

Also, the results of the survey complement the previous studies on investigating how the practice of DoD is applied \abc{and what DoD items are}, e.g., \cite{Silva2017}, \cite{Silva2018}. The respondents reported that there are multiple ways of creating and maintaining DoDs. Therefore, as future research, it would be worth investigating the advantages and disadvantages of different approaches. \abc{ Our analysis of 24 DoDs provided by the respondents showed that DoD items concern different subjects (like code, code reviews, tasks etc.) and their contents might be similar across different organizations. Thus, studying more DoDs might be valuable to broaden the knowledge about what DoD items are and how they are formulated. Possibly, it might additionally lead to learn how to efficiently and effectively specify DoD items. }

% e.g., various roles take care of DoD, counter-intuitively sometimes Developers do not establish DoD, sometimes DoD is established at the beginning of the project and another time in multiple rounds. Since there is no single way how to use DoD, it is worth investigating what the advantages and disadvantages of each approach are.

Moreover, new research directions are opened by two groups of problems---the problems related to the specification of DoDs and the cost of creating and maintaining DoD. It would be worth considering the possibility of supporting or even automating to some degree the process of creating DoDs. 

Finally, from the empirical software engineering perspective, we would like to draw the attention of other researchers to the usability of the questionnaire used in their survey, especially the presentation of the questions. It might have a significant impact on the response rate. It follows from the analysis of the drop-outs in our study that a significant number of our respondents left when they were presented a list of benefits or a list of problems to assess (i.e., a list of several items to assess on a scale with more than 5 values). It seems that it might be caused by the presentation method of the list-type of questions, e.g., the impression that answering the question would be overwhelming or time-consuming.  

%Since using DoD is an agile practice and DoD constitute of items which are requirements, it seems valuable to start the investigation by validating the already existing RE practices used in agile projects.

%co tu napisac?
%- jest wazna tak jak sugeruja inni [] i rozszerzamy o systamtycza identyfikacje benefitow
%- jak wynika z literatury istnieje malo zidentyfikowanych problemow, my rozszerzamy 
%i pokazujemy ze konkretne problemy wystepuja dosc czesto i moga miec spore znaczenie, stad glebsza analiza %bylaby bardzo ciekawa 
%- pokazujemy jak ludzie tworza i utrzymuja DoD. Rozszerzamy to co napisala Silva i wydaje sie ze nie ma %jednego sprawdzonego sposobu tak, ciekawe by bylo zbadanie ktora technika jest najbardziej efektywana, %dlaczego i w jakim kontekscie
%Our results suggest that 
%However, existing research 

\subsection{Implications for practitioners}
Our findings show that the DoD practice is perceived as useful by nearly all practitioners who used it. The primary benefits are about helping to make work items complete to ensure product quality and that the activities other than coding are executed, and all quality gates are passed to keep the product releasable. 

However, when deciding to use DoD or when thinking about improving that practice, one needs to understand that there might appear some problems along the way. The list of problems and their evaluation from the frequency and significance perspectives might help practitioners to assess the risks related to adopting the DoD practice and prepare in advance.

\abc{The results of our study showed that problems related to how DoD items are formulated correlate with nearly all other DoD-related problems (including those that could lower the chances of materializing benefits from using DoD). We believe that such problems can be easily mitigated by reviewing DoD documents. We propose a set of DoD quality criteria called VIRUPCUS that correspond to each of the aforementioned problems (\textbf{V}erifiable (P8), \textbf{I}dentified (P6), \textbf{R}elevent (P15), \textbf{U}p-to-date (P14), \textbf{P}recise (P5), \textbf{C}orrect (P18), \textbf{U}nambigious (P7),  \textbf{S}atisfiable (P19)). The criteria can be used as a checklist. For instance, when reviewing one of the DoD items provided by respondents stating that \emph{``burndown [ should be ] as close to deadline as possible},'' we could argue that it is \emph{imprecise} and impossible to \emph{verify}. Another example could be a DoD item stating that \emph{``automated test [ are ] written.''} We perceive this DoD item as \emph{incorrect} because we suspect} that the real intent behind formulating it was to have automated test cases that pass. Also, it is \emph{imprecise} because we do not know ``how many'' tests should be written. Finally, the word ``written'' seems \emph{ambiguous} in this context---does it mean implemented?

%The problems might be related to how a team create and use their DoD, e.g., high effort of creating a DoD or to verify if it is satisfied, and differences in understanding its items. Moreover, there might be some problems with formulating DoD like missing, unclear, or impossible to verify DoD. Unfortunately, those problems appear quite frequently and might significantly negatively impact a project. However, the proven techniques to solve the problems include mainly team cooperation (e.g., workshops, brainstorming sessions) which not only would improve the DoD practice, but might positively impact how team work together.  Sometimes, however, help from more knowledgeable persons like coach can be the solution.

It is also important to be aware of the fact that when a team dismisses to use DoD in their project, several important risks might materialize, such as, not meeting the deadline, lack of shared understanding of what ``done'' means, defects, low quality, and technical debt.

\abc{Moreover}, the results of our study show how practitioners establish and maintain DoD  which can be used for educational purposes. It might be especially useful for novice Scrum adepts who look for guidelines \abc{on how to create a DoD} or for those who seek some ideas on how to improve their process. Together with the information about the benefits, it might provide a strong argument in favor of using DoD for those who are reluctant (e.g., for Scrum Masters or agile coaches who would like to convince those undecided).

\abc{Finally, another educational application of the results of our study could be for those looking for what their DoD might contain or what they can add to their DoD. Those practitioners can look into the categories or the subjects of DoD items to support their, e.g., brainstorming session in their team. Additionally, the ten statements that can be used to formulate DoD items might also trigger a discussion of whether they are relevant in their context or not.}

%Therefore, these observations seem to confirm the observations made while analyzing the \emph{Significance} measure that the problems related to how DoD is formulated are of big importance.

\section{Conclusions}
\label{sec:conclusions}

The goal of the study presented in this paper was to investigate the usefulness of using the Definition of Done (DoD) practice. To achieve that goal, we conducted a questionnaire-based survey. In the research study 137 practitioners from all over the world took part.

Within the study, we looked at the usefulness of DoD from two perspectives: the benefits it might bring to a project or product team, and the problems it might cause. Additionally, we investigated how DoDs are created and maintained to extend the existing body of knowledge.

Our findings show that the vast majority of practitioners (93\%) perceive DoD as at least valuable and confirm that its presence in the project help to make work items complete, assure product quality, ensure the needed activities are executed, ensure that all quality gates are passed, and help to keep product releasable. The usefulness of DoD can be also confirmed by an analysis of problems that might appear when DoD is not used. Our respondents mentioned defects, low quality, technical debt, lack of shared understanding of what done means, and not meeting the deadline as those most frequent.

On the other hand, the results of the survey show that there are some common problems related to using DoDs, and some of them might even significantly impact an agile project (e.g. high effort of implementing and using DoDs or different understanding of DoD items).

Moreover, the most important problems with using DoDs relate to imprecise, missing, or unclear DoD items. These problems are getting solved by agile practitioners mainly internally within a team through some collaborative activities like workshops or, in some cases, with the help of an agile coach. \abc{We proposed a set of DoD quality criteria (VIRUPCUS) that might be used as a tool to support reviewing DoD documents.}

Furthermore, the respondents characterized the process of creating \abc{and using a DoD and what DoD items are}. Although there are some variations in that process, the most general steps are: (1) propose a DoD, (2) analyze it, (3) adjust the proposal, (4) document, sign off and publish the DoD, (5) use it, (6) evaluate, and improve the DoD. In most cases, the DoD was created by the members of Scrum Team, however, we observed that in nearly 22\% of the studied projects, the developers were not involved in that process. \abc{We also found that most DoD items concern quality management and process management, and are about tests, code reviews, documentation and acceptance criteria. The subjects and contents of some DoD items across different companies is the same or similar with some variability.}

The findings from the study allowed us to identify new research directions, i.e., the necessity of in-depth analysis of the DoD practice, of the identification of the means of preventing or mitigating the problems related to using DoDs, and, finally, of the investigation of the ways to maximize the benefits that come from using DoDs.

\section*{Acknowledgment}
The authors would like to thank all participants of the study for sharing their experience and opinions.

\section*{References}

\bibliographystyle{elsarticle-num}
\bibliography{references}

\end{document}